\documentclass[10pt]{article}
\usepackage[a4paper, tmargin=0.75in, lmargin=0.75in, rmargin=0.75in, bmargin=0.75in]{geometry}
\usepackage{xcolor}
\usepackage{csquotes}
\usepackage{algorithm}
\usepackage{algpseudocode}

\usepackage[style=authoryear,backend=bibtex,maxcitenames=2,maxbibnames=99]{biblatex}
\bibliography{biblio}
\DeclareNameAlias{sortname}{last-first}
\makeatletter
\AtEveryBibitem{%
  \global\undef\bbx@lasthash%
  \clearfield{}}
\makeatother

\usepackage{authblk}
\usepackage{fullpage}
\usepackage{enumitem}
\usepackage{framed,color}

\usepackage{float}%
\floatstyle{plaintop}%
\restylefloat{table}

\usepackage{mathrsfs}
\usepackage{amsmath}
\usepackage{amsthm}

\usepackage{accents}
\usepackage{amssymb}

\usepackage{graphicx}
\graphicspath{{Images/}} 
\usepackage{eso-pic} 
\usepackage{subfig} 
\usepackage{caption} 
\usepackage{transparent}
\usepackage{booktabs} 

\usepackage{url}
\usepackage{breakurl}
\usepackage{hyperref}%
\hypersetup{
colorlinks=true,
    linkcolor=blue,
    filecolor=magenta,      
    urlcolor=blue,
    citecolor=blue
}
\date{}
\title{\textbf{A dynamic latent space time series model to assess the spread of mumps in England}}
\author[1,2]{Hardeep Kaur}
\author[1]{Riccardo Rastelli}
\affil[1]{\footnotesize School of Mathematics and Statistics, University College Dublin, Dublin, Ireland;}
\affil[2]{\footnotesize Insight: Centre for Data Analytics, University College Dublin, Ireland;}

\begin{document}
	
	\maketitle
\begin{abstract}
\noindent

This work is motivated by an original dataset of reported mumps cases across nine regions of England, and focuses on the modeling of temporal dynamics and time-varying dependency patterns between the observed time series. The goal is to discover the possible presence of latent routes of contagion that go beyond the geographical locations of the regions, and instead may be explained through other non directly observable socio-economic factors. We build upon the recent statistics literature and extend the existing count time series network models by adopting a time-varying latent distance network model. This approach can efficiently capture across-series and across-time dependencies, which are both not directly observed from the data. We adopt a Bayesian hierarchical framework and perform parameter estimation using L-BFGS optimization and Hamiltonian Monte Carlo. We demonstrate with several simulation experiments that the model parameters can be accurately estimated under a variety of realistic dependency settings. Our real data application on mumps cases leads to a detailed view of some possible contagion routes. A critical advantage of our methodology is that it permits clear and interpretable visualizations of the complex relations between the time series and how these relations may evolve over time. The geometric nature of the latent embedding provides useful model based summaries. In particular, we show how to extract a measure of contraction of the inferred latent space, which can be interpreted as an overall risk for the escalation of contagion, at each point in time. Ultimately, the results highlight some possible critical transmission pathways and the role of key regions in driving infection dynamics, offering valuable perspectives that may be considered when designing public health strategies.\\


\end{abstract}

\section{Introduction} \label{Introduction}

Mumps is a viral infection that is transmitted primarily through respiratory droplets, similarly to colds and flu. Mumps may spread when infected droplets of saliva are released into the air and are either inhaled or transferred from contaminated surfaces to the mouth or nose. Mumps typically affects children aged 4 to 16 and young adults who did not receive the MMR (Measles, Mumps, and Rubella) vaccine during their childhood vaccination schedule. In vaccinated individuals, mumps tends to present with milder symptoms and fewer complications. The MMR vaccine was introduced in England in 1988, but also people born in the late 1990s and early 2000s who may have missed out on vaccination during that period remain at higher risk. 

Outbreaks of mumps in England tend to occur in settings where individuals live or work in close contact, amplifying the risk of the spread. The most notable outbreak in recent history took place in 2005, primarily caused by a shortage of MMR vaccines, leading to an unusually high percentage of the population remaining unvaccinated.
Outbreaks can lead to spillovers between different geographical areas, potentially leading to local or global epidemics. 
For this reason, it is essential that the risk of contagion (and spillover) is carefully assessed and taken into account when making decisions on health policies and interventions.
While the geography of England definitely plays a role in determining the spread of disease, it would be naive to consider this as the only relevant factor to take into account. Indeed, as mentioned, uptake of vaccines and other socio-economical factors also play a key role in determining the prevalence and spread of the disease. Social networks, professional networks, and mobility networks connect individuals in a number of non-trivial ways, which are very difficult to observe and objectively measure. This suggests that a modeling framework based on latent factors may represent a viable choice to explain the data and its complex patterns, without necessarily relying on a large number of covariates that can only be used as proxies at best.

Building on the recognition of these complex, interrelated factors influencing the spread of mumps, our study focuses on the analysis of reported mumps cases across nine regions of England, specifically examining the period from 2008 to 2020. To address the challenges posed by the intricate transmission patterns and latent factors at play, we introduce an original time series model that combines a log-linear time series model and latent position network models. Our final goal is to analyze the temporal dynamics and transmission patterns of mumps, providing a detailed view of how the disease has been spreading across different regions. This framework also allows us to explore broader statistical relationships between the regions, treating the spread of disease as part of a larger, interconnected system.

From the statistical methodological point of view, this paper makes a number of contributions which advance the recent literature on network count time series modeling. Multivariate time series may be seen under the lenses of network analysis, with time series being represented as the nodes of a network. The connections between the nodes of the network, represented as edges between the nodes, describe instead the pairwise relationships among the time series. Network-based time series models have become a popular approach for analyzing multivariate data, since they provide a structured framework to uncover underlying patterns and dependencies. The network framework permits a more parsimonious model specification, where the additional structured layers tend to favor a more efficient use of model parameters. Critically, the embedding of data permits a more clear and interpretable visualization of the data, which can lead to new perspectives and results.

A point to note regarding our contribution is that we take care of the network framework using both static and dynamic models. The temporal feature can be essential for a statistical model in order to capture the inherent dynamic behavior of the time series. Time-evolving networks, unlike their static counterparts, offer a more robust way to understand the associations between variables by reflecting their true relationships as they change over time. Thus, we directly extend our previous framework (\cite{Kaur_rastelli_2024}) and present here a development into studying dynamic time series network models, with a focus on the analysis of mumps transmission patterns in England.

Similarly to our previous work (\cite{Kaur_rastelli_2024}), we employ the Latent Position network Model (LPM) to give a structure to the interactions between the time series. A key novelty of this paper is that we focus on the distance model, which uncovers the strength of connections between the nodes based on the Euclidean distance that separates them. The advantage of the distance model, compared to the projection model used in \textcite{Kaur_rastelli_2024}, is that the distance model can give more interpretable visualizations, while maintaining a comparable flexibility and inferential setup.

In fact, we illustrate via various simulated experiments that our analysis method can tackle and accurately infer a number of complex patterns from the data, thus being able to represent a multitude of realistic scenarios. As regards the output of our modeling, the geometric nature of our LPM permits the creation of model-based summaries that can illustrate some new perspectives about the dataset. In the context of our original mumps dataset, we introduce a measure of expansion and contraction of the latent space: this has an immediate and straightforward interpretation since it provides an aggregate and model-based measure of the possibility of spillovers and contagion. Thus, this model can be used to quantify the evolution of epidemic risk and, ultimately, it can be helpful in making critical decisions regarding vaccination programs and other related medical interventions.

We emphasize that our model hinges on a latent variable specification, whereby both the time series interactions and the underlying network structure are not observed, unknown, and completely inferred from the data. As such, the role of the latent variable is to capture not only the relevance of some information that may potentially be observed (e.g. the geographical location of the nodes); but also capture many other non-observed, non-observable, and perhaps abstract socio-economic factors that possibly play a central role for this dataset.

\subsection{Related Work}

The study of multivariate time series models using network approaches has attracted considerable attention over the last few years. The traditional time series models, such as one-dimensional autoregressive models (\cite{box2015time}), can fall short in capturing the dependencies between the time series. Vector Autoregressive (VAR) models (\cite{book_Luetkepohl_2005}), provide a very flexible structure and are largely used in econometrics and other research fields. However, these models can often be over-parametrized, especially when considering model orders larger than one. This serves as a primary motivation to consider VAR extensions using networks, which enforce more structure into the model, thus promoting parsimony.

Other recent related works include that of \textcite{knight2016modelling}, who introduced the Network AutoRegressive (Integrated) Moving Average (NARIMA) models, which define multivariate continuous time series coupled with a network structure. Their model successfully tracked the dynamics of mumps outbreaks in UK in 2005 by directly modeling the network of interactions based on the movements of infected individuals between different areas of the UK using NARIMA processes. While our approach shares similarities with \textcite{knight2016modelling}, as both works involve network time series to model the spread of mumps, our method differs by assuming that the network structure is not known and evolves dynamically over time.
\textcite{zhu2017network} introduce instead the Network Vector AutoRegressive (NAR) model, an extension of the traditional VAR model, which incorporates a structured network framework. In this model, a node's value is influenced not only by its own history, but also by the average historical values of its neighboring nodes. The NAR model also includes node-specific covariates, reducing complexity by using a constrained set of parameters, which improves both manageability and interpretability. Inference is performed using least squares in two asymptotic settings: (a) increasing the time sample size ($T \rightarrow \infty$) with a fixed number of nodes $N$, and (b) both $N$ and $T$ increasing simultaneously. \textcite{JSSv096i05} introduce the Generalized Network Autoregressive (GNAR) model for continuous random variables, building on the model proposed by \textcite{zhu2017network}. This extension incorporates node-specific effects based on the sizes of neighboring networks. \textcite{Chen2020CommunityNA} propose the community network vector autoregressive model, enhancing the flexibility and generality of model formulated by \textcite{zhu2017network} by allowing heterogeneous effects across different network communities. \textcite{nason2024modellingclustersnetworktime} presents a community-$\alpha$ GNAR model that extends the GNAR framework, designed to identify and model community structures within network time series, which can reveal interesting dynamics, especially in the context of presidential elections in the USA. This model is particularly useful for high-dimensional settings. Our work extends the foundational study by \textcite{zhu2017network} and its extensions on network time series by modeling count time series with an unknown, dynamic network structure, unlike the aforementioned models that assume a known fixed network and continuous data.

A related strand of literature focuses on continuous time series data where, similar to our approach, the network structure is unknown. However, unlike our work, these studies typically assume a fixed network structure rather than modeling it changes over time. The work of \cite{Bolstad_2011} addresses the problem of inferring sparse causal networks using multivariate autoregressive processes, by establishing conditions under which a group-lasso method can consistently estimate network structures. \cite{basu_2015} introduces Granger causal models to learn a network structure from temporal panel data. The authors introduce a group-lasso regression regularization framework to facilitate this learning process. Additionally, they explore a threshold-based variant to tackle issues related to group misspecification. \textcite{barigozzi2019nets} introduce a network-based approach for analyzing multivariate time series, defining a network for multivariate time series as a graph where the vertices represent the components of the time series, and the edges are inferred through the use of the long-run partial correlation matrix between multiple time series. \textcite{ahelegbey2017bayesian} presents a Bayesian hierarchical model combining a VAR model, Covariance Graphical Model (CGM), and a LPM model, their work centers on continuous financial time series. Additionally, their approach employs a CGM to model the network and an LPM framework to uncover the spatial position of the nodes.

Although there is extensive literature on modeling and inference for time series with continuous responses, research on multivariate count time series models for network data remains limited. Recently, work by \textcite{armillotta2023} has significantly contributed to this area. They expanded the Network Autoregressive Model (NAR) to specifically accommodate count data, addressing the complexities associated with non-random neighborhood structures. This extension resulted in the development of the Poisson Network AutoRegression (PNAR) model. \textcite{armillotta2023} detail the PNAR(p) model specification for both linear and log-linear cases, including discussions on stability properties, which are essential for understanding the model's behavior. Their research comprehensively examines two asymptotic regimes discussed in \textcite{zhu2017network} and proposes a robust theoretical framework for asymptotic inference. In their work, \textcite{amillotta2022generalized} introduce a statistical framework that unifies the findings of \textcite{zhu2017network} and \textcite{armillotta2023}. This framework is designed to handle both continuous and count responses over time for each node within a known network, showcasing its practical applicability. \textcite{liu2023newmethodsnetworkcount} reviews existing count time series models and presents two new models, GNARI (Generalized Network Autoregressive Integrated) and NGNAR (Negative Generalized Network Autoregressive) for count network time series.
\textcite{yin2024functionalcoefficientsnetworkautoregressive} presents the Functional Coefficients Network Autoregressive (FCNAR) model, which extends existing count time series network models by incorporating both nonlinear autoregressive and network effects in a multivariate context. This model allows for a more flexible approach to understanding complex relationships within data sets. The above studies assume a fixed network structure across time. In contrast, our work contributes to the literature on network count time series by introducing a time series modeling framework inspired by the work of \textcite{armillotta2023}, with the key distinction that, in our setting, the network structure is unknown and evolves dynamically over time.

Another strand of literature that is related to our work relates to networks that evolve over time. In the dynamic networks literature, various models have been developed to extend models from the static network literature. The dynamic extension of the Latent Position Model (LPM) (\cite{Hoff20021090}) can be found in a number of works including \textcite{Sarkar20051145}, \textcite{Sewell20151646}, \textcite{Friel20166629}, \textcite{Durante20162203} and, \textcite{rastelli2021continuous}. Other related dynamic networks models are extensions of the stochastic blockmodel, by \textcite{Matias_Miele_2017}, \textcite{Ludkin_Eckley_Neal_2018}, and \textcite{Pensky_2019}. 

More recently, some works have considered dynamic network models to study the dependency structure in multivariate time series. \textcite{kang2017dynamic} propose a network autoregressive model that accounts for a time-varying network structure by combining multi-scale modeling with network-based neighborhood selection. The model is designed to capture local temporal structures and detect significant changes in the network over time, all while ensuring sparsity in the interactions. \textcite{Krampe_2019} further extents the work of \textcite{knight2016modelling} and \textcite{zhu2017network} by modeling random network structure. \textcite{Krampe_2019} introduces a multivariate doubly stochastic time series framework that allows for the separate modeling of dynamic attributes and the underlying network structure. 
The paper describes a vertex-labeled dynamic network represented by a weighted time-dependent adjacency matrix. The methodological framework of \textcite{zhu2017network} is expanded by \textcite{JSSv096i05}, who propose an extension of the GNAR model, by accommodating time variations in the network structure using time-dependent weights and neighborhood structures. Another relevant contribution in this context is of \textcite{CASTRO2012253} who study count time series using embedded network structure. The authors combine spatial and temporal dependencies within the count data modeling. This work effectively incorporates spatial dependencies by applying a spatial structure to the latent continuous variables. The framework also accommodates both time-stationary and time-varying temporal correlation patterns. \textcite{Tjøstheim_Jullum_Løland_review_2023} review of some recent developments in the literature of time series and dynamic networks, noting that the literature on time series with dynamic network embeddings is much more sparse than the static case.

Our work builds on LPMs, popularized by \textcite{Hoff20021090} as a statistical model for social networks. This highly influential work resulted in a significant increase in the research on LPMs, yet, to our knowledge, no studies have applied LPMs to uncover temporal network structures in multivariate count time series. This gap in the literature is a key focus of our contribution to network time series research.

The rest of the paper is organized as follows: Section \ref{dynamic_tslpm} describes our proposed model. Section \ref{Inferential Procedure} outlines the inferential process. Section \ref{Simulation study} contains simulation studies exploring the performance of our methodology on multivariate count time series with underlying non-random and random network structures. Section \ref{real_dataset} includes the application of the proposed model using the original dataset on mumps cases in England. Section \ref{discussion} concludes the paper with a conclusion and discussion on the possible extensions.

\section{Dynamic Time series Latent Position model (DTSLPM)}
\label{dynamic_tslpm}

\subsection{Data}
The observed dataset consists of a multivariate time series $\mathbf{y}_{t}=\{y_{it}, t=1,2, \dots, T$ and $i=1,2, \dots ,N$\}, for $N$ different time series and $T$ time-points. We assume that the data is Poisson distributed, and let $ \boldsymbol{\lambda}_{t} =\{(\lambda_{it}) , t=1,2, \dots ,T$ and $i=1,2, \dots ,N$\} be the corresponding intensity process. Define $\mathscr{F}_{t}$ be the $\sigma$-field generated by $\mathbf{y}_{t-1}$. According to the model specification, we make the assumption that $\lambda_t = E(\mathbf{y}_{t}|\mathscr{F}_{t-1})$. A network time series ${\cal X} = (\mathbf{y}_{t},{\cal G})$
is a stochastic process composed of a multivariate time series $\mathbf{y}_{t}$ and an underlying network ${\cal G}=({\cal K},\mathcal{E})$ where ${\cal K}=\{1, \dots , N\}$ is the set of nodes on which time series are observed, $\mathcal{E} \subset {\cal K} \times {\cal K}$ is the set of weighted edges, and ${\cal G}$ is an undirected graph with $N$ nodes. This article focuses on the situation where the $y_{it}$ multivariate time series are counts, that is integers greater than or equal to zero accounting for a time-varying network structure with a fixed number of nodes and varying edge weights varying across time denoted as $\mathcal{E}_{t}$.

\subsection{Model Specification}

We present a multivariate time series framework characterized by a log-linear autoregressive model structure for count data, popularized through a number of works which include \parencite{Doukhan2017MultivariateCA, FOKIANOS2021, Davis_Fokianos_Holan_Joe_Livsey_Lund_Pipiras_Ravishanker_2021}. The proposed model exhibits a hierarchical structure and we name it the Dynamic Time Series Latent Position Model (DTSLPM) of order 1, defined as follows:
\begin{gather} 
y_{it}|\mathscr{F}_{t-1} \sim Pois(\lambda_{it}) \notag \\
\log(\lambda_{it}) = \alpha + \sum_{j} \gamma_{ij(t-1)} \log(y_{j(t-1)}+1) \notag \\
 \gamma_{ijt}=   \begin{cases} 
      \beta_{i}  &  i=j \\
      \frac{2   }{1+\exp(d_{ijt})}  & i \neq j \\
   \end{cases}
   \label{eq:DTSLPM_model}
\end{gather}
where, $d_{ijt} = d(\textbf{z}_{it},\textbf{z}_{jt}) =\sqrt{\sum_{i=1}^{K} (z_{itk} -z_{jtk})^2}$ for $i,j = 1,\dots,N$ and t= $1, \dots T$. The parameters of the model that need to be estimated are  $\alpha \in \mathbb{R}$, $\boldsymbol{\beta} = \{\beta_1, \dots, \beta_N\} \in \mathbb{R}^N$, $\boldsymbol{\cal {Z}}_{t} = \{\mathbf{z}_{it} \in \mathbb{R}^2| i = 1, \dots, N\ \text{and} ~ t= 1, \dots T\}$. In the proposed model, the latent space that we consider is two-dimensional so that we facilitate the visualization of network structure. 

The model states that the log rate of the mean of the $i$-th series at time $t$ is influenced by the log values of the $i$-th series itself, using an autoregressive framework. The effect of this term is quantified by the parameter $\beta_i$. In addition, the $i$-th series is also driven by the log values of the other series, each weighted by the coefficient $\gamma_{ijt}$. The presence of the unit value within the logs is to facilitate numerical stability. The parameters of the model $\boldsymbol{\beta}$ and $\boldsymbol{\gamma}$ may be zero or any positive or negative value, and their interpretation is analogous to the coefficients of a VAR time series model. 

The association between any two series at any time-step $t$ is described by the interaction parameter $\gamma_{ijt}$ through a logistic function on $d_{ijt}$ (\cite{Sarkar20051145}). This choice is motivated by the fact that it enables the quantification of the connection weight based on distances with a fairly simple interpretation. Whenever two nodes are close to each other, the interaction term has a higher value, capped at $1$, which is attained when the two nodes have exactly the same position. Imposing this upper bound is reasonable since higher values would yield a non-stationary process, as discussed later in Section \ref{Stability_condtn}. 
If two nodes are far from each other, then the interaction term gets small, tending to zero when the distance increases. This can be interpreted as a situation of independence between the time series. We note that negative values of the interaction terms cannot be obtained: this is a model assumption that we make, since negative interactions typically lead to less clear interpretations, and because they are difficult to represent with the geometric nature of our model based on pairwise distances. 

The parameter $\alpha$ is the intercept of the model and it is constant over time. This facilitates the convergence of the series to a non-zero mean process. The series-specific autoregressive parameter $\beta_i$ is also constant over time, and its interpretation is equivalent to that of a autoregressive model parameter: a positive value of $\beta$ leads to sustained trends and positive autocorrelations, whereas a negative value leads to more erratic behavior, unclear trends, and, possibly, alternating signs on the autocorrelation plots. More information on the interpretation on of the model parameters can be found in the previous work \textcite{Kaur_rastelli_2024}.

The latent positions $\boldsymbol{\cal {Z}}_{t}$ are a central feature of interest for our model. Since this latent space evolves over time, the latent nodes are characterized by a latent trajectory across the Euclidean space. We assume that, apriori, these latent trajectories are random walks, thus, similarly to \textcite{Sarkar20051145,Friel20166629,Sewell20151646,Sewell2016105}, we specify a Markov process characterized by the following initial distribution:
\begin{equation} 
\pi({\boldsymbol{\cal Z}}_1|\boldsymbol\phi)= \prod_{i=1}^{N} {\cal{N}} ({\bf z}_{i1}|{\bf 0},\rho^{2} \mathbf{I}),
\end{equation}
and transition equation:
\begin{equation}
\pi({\boldsymbol{\cal Z}}_t|{\boldsymbol{\cal Z}}_{t-1},\boldsymbol\phi)=\prod_{i=1}^{N} {\cal{N}} ({\bf z}_{it}|{\bf z}_{i(t-1)},\sigma^{2} \mathbf{I})
\label{transition}
\end{equation}
for $t=2,3,\ldots,T$.  For a latent space $\mathbb{R}^2$,  ${\bf z}_{it}$ is the vector of the $i^{th}$ actor's latent position at time $t$, and $\boldsymbol{{\cal Z}}_t$ is the $N\times 2$ matrix whose $i^{th}$ row is ${\bf z}_{it}$, where $\mathbf{I}$ is the $2\times 2$ identity matrix, $N({\bf x}|\boldsymbol{ \mu},\boldsymbol{\Sigma})$ denotes the normal probability density function with mean $\boldsymbol\mu$ and covariance matrix $\boldsymbol{\Sigma}$ evaluated at ${\bf x}$, and $\boldsymbol\phi$ is a vector of parameters.

In the definition of our model, we achieved the desired property that the strength interaction between two nodes $i$ and $j$ at time $t$ increases as the distance between their respective latent positions $\mathbf{z}_{it}$ and $\mathbf{z}_{jt}$ in the latent space decrease. Consequently, the underlying networks $G = ({\cal K},\mathcal{E})$ that describe the conditional dependencies among the time series across the system at different time points are determined by the estimated weighted interconnections. In order to preserve the information across successive time steps, we make a key assumption within this study: we assume that a node can transition within latent space across time intervals, but significant transitions are improbable. This is achieved by setting the parameter $\sigma^2$ to a small value. This choice facilitates more clear visualizations since the movement of nodes tends to be the minimum movement that is required to explain the data. Besides, there could be an argument to bring up if large transitions are allowed, in that the model will try to overfit the data by having fewer constraints on its temporal evolution. 

Finally, we briefly note that in this paper we arbitrarily fix the number of dimensions (K) in the latent space to $2$. This decision is especially common in the literature on LPMs, albeit not necessarily well founded from a purely statistical perspective. Indeed, a clear goal would be to try to infer, perhaps automatically, the ideal number of dimensions from the given dataset. We do not pursue this goal here, as we feel that this is a problem that is currently being tackled by the literature on LPMs, and for which we do not have a clear solution yet. So, we make a pragmatic decision of choosing two latent dimensions, since this allows for very clear representations in $\mathbb{R}^2$, while not compromising the flexibility of the model or its nature.

\subsection{Stability of network time series process} \label{Stability_condtn}

To ensure the stability of the network time series process denoted by the pair ${\cal X} = (\mathbf{y}_{t},{\cal G})$, it is critical to establish the stability of the multivariate count time series $\mathbf{y}_{t}$ modeled using log-linear model which takes the form:
\begin{equation}
    \mathbf{y}_{t} \sim \text{Poisson}(\boldsymbol{\lambda}_{t})  ~~~ \log(\boldsymbol{\lambda}_{t}) = \alpha + \Gamma \log(\mathbf{y}_{t-1} + \mathbf{1}_{N}) 
\end{equation}

Building on the discussion by \textcite{Doukhan2017MultivariateCA}, and further elaborated in our previous work (\cite{Kaur_rastelli_2024}), a key sufficient condition required to ensure the stability and stationarity of the joint process  $(\mathbf{y}_{t},\boldsymbol{\lambda}_{t})$ is:
\begin{equation*}
-1 < \min_{i}(\beta_{ii} - r_{i}) < |\Omega_{\text{max}}(\Gamma)| < \max_{i}(\beta_{ii} +r_{i}) < 1
\end{equation*}
where $\Omega_i(\Gamma)$ denotes the eigenvalues of $N \times N$ matrix $\boldsymbol{\Gamma}$, $\beta_{ii}$ is the diagonal entry in row $i$, and $r_{i} = \sum_{i \neq j} |\gamma_{ij}|$ is the sum of the absolute values of the off-diagonal elements in row $i$ of $\boldsymbol{\Gamma}$.
The above inequality gives a sufficient condition on the interaction matrix of the model to ensure the stationarity of the process. More details can be found in \textcite{Kaur_rastelli_2024}.

\subsection{Prior specification} \label{Prior_specification}
We adopt a Bayesian framework to estimate the model parameters. In particular, we allocate independent and non-informative priors to the parameters $\alpha$, $\boldsymbol{\beta}$, and $\boldsymbol{{\cal Z}}_t$. Both $\alpha$ and $\boldsymbol{\beta}s$ are assigned independent normal priors with a mean of zero and a standard deviation of 10. The choice of a large standard deviation is intentional, as it reduces the impact of prior assumptions on the posterior distributions. 

As mentioned, we assign independent Gaussian priors to the increments of the parameters $\boldsymbol{{\cal Z}}_t$. The initial latent positions $\boldsymbol{{\cal Z}}_1$ arise from a zero mean Gaussian prior with high standard deviation $\rho$. 
At each time frame $t$, the set of two-dimensional latent coordinate matrix $\boldsymbol{{\cal Z}}_t$ are generated using the positions at the previous time frame $\boldsymbol{{\cal Z}}_{t-1}$ as a mean, and a small standard deviation $\sigma$ is used to penalize large displacements over time (\cite{Sarkar20051145}). 
The parameters $\rho$ and $\sigma$ may be either be gamma distributed with shape and rate hyperparameters \{$a_{\rho}, b_{\rho}, a_{\sigma}, b_{\sigma}$\}, respectively; or they may be considered hyperparameters themselves and thus be arbitrarily set. A regards the gamma prior approach, we follow previous works which include \textcite{Sewell20151646,Sewell2016105,Friel20166629}. In those applications where we fix the values of these standard deviations, we set them as $\rho = 10$ and $\sigma = 0.05$.

\section{Model Fitting} \label{Inferential Procedure}
\subsection{Estimation}

We consider two primary methodologies for inference on the parameters of a DTSLPM, implemented through the STAN framework (\cite{STAN_SOFTWARE}).
One of these methodologies uses the Limited-memory Broyden-Fletcher-Goldfarb-Shanno (L-BFGS) algorithm (\cite{NoceWrig06}) to optimize the likelihood function, which is particularly effective for large-scale problems like DTSLPM. The algorithm iteratively updates parameter estimates by combining information about gradients and curvature, efficiently converging even when the number of parameters increases with the size of the time series.

In addition to the optimization-based approach, we also employed Hamiltonian Monte Carlo (HMC) \parencite{1987PhLB..195..216D, Neal2011MCMCUH, betancourt2013hamiltonian}, a sophisticated Markov Chain Monte Carlo (MCMC) technique that uses Hamiltonian dynamics to sample from the posterior distribution of model parameters. In HMC, the model parameters represent the position in a high-dimensional space, while auxiliary momentum variables, typically following a multivariate normal distribution, are introduced to aid in sampling. This combination of position and momentum allows the algorithm to explore the parameter space more efficiently by leveraging gradients of the log-density function, facilitating better transitions, and reducing random walk behavior. 

Following standard practices of Bayesian model fitting, we primarily rely on the sampling approaches to fit our models, and run the sampling until we observe satisfactory convergence to a stationary distribution through trace plots and other related convergence diagnostics that are available in STAN.

\subsection{Post-processing of posterior samples} \label{Procrustes}

A key challenge in fitting Bayesian LPMs is that the likelihood function associated to the model will only depend on the latent positions through the pairwise distances. This means that the model is not identifiable up to any transformation of the latent space that preserves the pairwise distances between nodes. These transformations can be rotations, reflections, or translations of all positions simultaneously for all time frames. While this may not be necessarily an issue under an optimization setting, under a Bayesian context the circumstances become more problematic. The samples that we use to calculate posterior summaries are sensitive to rotations, reflections, and translations of the latent space, and thus should be suitably transformed before the aggregations. 

For this, we post process the posterior samples of latent locations employing the so-called Procrustes matching (\cite{Procrustes_2004}), adjusting the procedure of \textcite{Hoff20021090} to a dynamic network setting, as already done by \textcite{Friel20166629}. 
Since we are interested in transforming the complete trajectories, we stack the $N$ latent positions for each of the $T$ times frames, thus obtaining latent spaces made of $TN$ points for all nodes at all times. We then proceed with standard Procrustes analysis on these stacked matrices.
As a reference set, we take the Maximum-a-Posteriori (MAP) latent positions $\boldsymbol{\hat{ {\cal Z}}}$ which is a $(TN) \times 2$ matrix.  Then, for each latent space $\boldsymbol{ {\cal Z}}^{(1)}, \dots, \boldsymbol{ {\cal Z}}^{(M)} $ that we obtain during the sampling, we perform a Procrustes transformation on each of these matrices \parencite{Friel20166629, Sewell20151646,Sewell2016105} using the MAP as a reference. As a final operation, we un-stack each of the matrices used to obtain the transformed collection of points for every $N$, $T$ and sample.

\section{Simulation study} \label{Simulation study}
We examine the performance of our DTSLPM framework in the estimation of the model parameters, using a number of realistic scenarios where we try to mimic some typical dependency structures that are observed in real applications. 

\subsection{Static time series distance model} \label{static_DTSLPM}
Initially, we consider a simpler version of our model whereby the latent positions remain constant over time. As a consequence, the interactions between series do not change over time, and so the model may be defined as ``static''.
In the static version of the DTSLPM model, the form remains the same as in Eq.~\ref{eq:DTSLPM_model}, with the only difference being that $\gamma_{ij}$ has no time dependence. 

We conduct a simulation study across different network dimensions with varying sample sizes. The data generation process involves the simulation of the time series using three main parameters: the intercept parameter $\alpha$, the autoregressive coefficient for each series $\boldsymbol{\beta}$, and the latent positions $\boldsymbol{\cal Z}$.
The parameter $\alpha$ is generated from a uniform distribution between -3 and 3. In a similar manner, the parameter $\beta$ is also generated from a uniform distribution, ranging from -1 to 1. We note that the value of $\alpha$ does not have any meaningful effect on the stationarity of the process, whereas the autoregressive parameter necessarily has to be between $-1$ and $1$. 

In this section, the latent positions remain fixed over time, denoted as, $\mathbf{z}_{1}, \dots, \mathbf{z}_{N}$, and are generated independently from a multivariate normal distribution, ensuring that the series are organized into at least two clusters. The rationale behind the presence of clustering is that we aim to obtain interaction values that are appreciably different, from very strong to weak, which can lead to reasonable and realistic settings. If all nodes are positioned too far apart in the latent space, they may become independent from other nodes, making the estimation of their positions particularly challenging. By contrast, if all nodes are close to each then we obtain too many strong positive interactions and unstable behavior. The clustering framework represents then some middle ground that that can correspond to a realistic setting, with both dependence and independence patterns being present.

Each cluster is centered in the latent space with a unique mean vector, which determines its location, and all clusters share the same covariance matrix. Initially, the covariance matrix is diagonal with a variance of 0.1, which leads to fairly strong interactions for nodes that are in the same cluster. However, our goal is to obtain datasets whereby the interactions between series are meaningful, so that we can accurately estimate the interaction parameters and, as a consequence, the latent positions. For this reason, once an initial set of latent positions is created, we iteratively adjust the latent space by expanding the latent positions by a small factor. This iterative process continues recursively as long as the stability conditions are satisfied as outlined in Section \ref{Stability_condtn}. 
Through this expansion procedure, we ensure that the stability conditions for the time series remain satisfied, while at the same time we can obtain a compelling dataset with meaningful interactions.

Additionally, the number of clusters increases with the number of series. This choice is driven by the need to maintain stable series, as having too many nodes within a single cluster increases the likelihood that the interaction matrix will fail to meet the stability conditions. In consideration of this, we selected 2 clusters for datasets of 5 to 10 nodes, and as the number of nodes increased to 15, we increased the number of clusters to 3.

\subsection{Large scale simulation study using optimization}  

We consider datasets with $N = {5, 10, 15}$ and time frames $T = {100, 500, 1000}$. Specifically, we simulate multivariate time series datasets with $N = {5, 10}$ in two clusters located at $(0,0)$ and $(3,3)$, and for $N = 15$, we use three clusters located at $(0,0)$, $(-1.8, 3)$, and $(-3, -3)$ in the latent space. This choice is driven by the need to maintain stable series, as having too many nodes within a single cluster increases the likelihood that the interaction matrix will fail to meet the stability conditions. 
All covariance matrices are defined as $\Sigma = diag(0.1,0.1)$, where this choice of mean vectors and covariance matrix leads to strong within-cluster connections and weak between-cluster connections.
For each different pairs of values of $N$ and $T$, we generate 100 different realizations of multivariate time series data with varying random seeds. We employ the L-BFGS optimization on each dataset to obtain point estimates by maximizing the joint posterior distribution from the model. Then, we compare our estimated parameters with the true parameter values that have generated the data.

\begin{figure}[htbp]
    \centering    
    \includegraphics[width = 0.496 \textwidth]{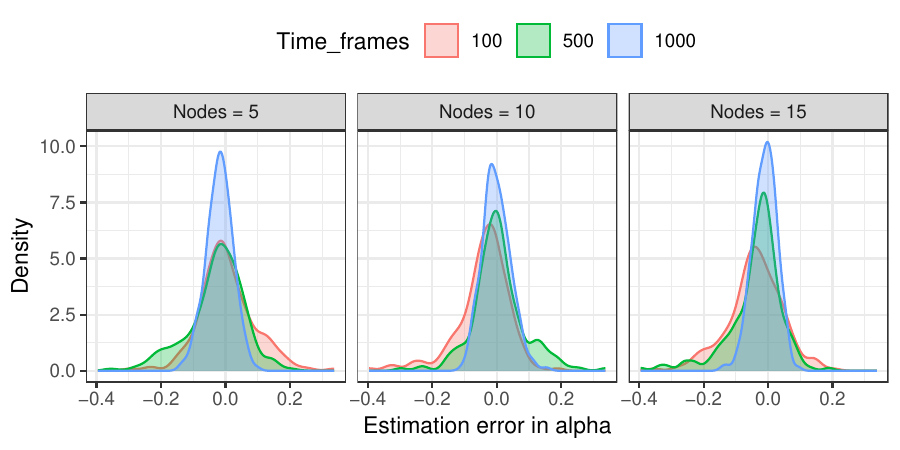}
    \includegraphics[width = 0.496 \textwidth]{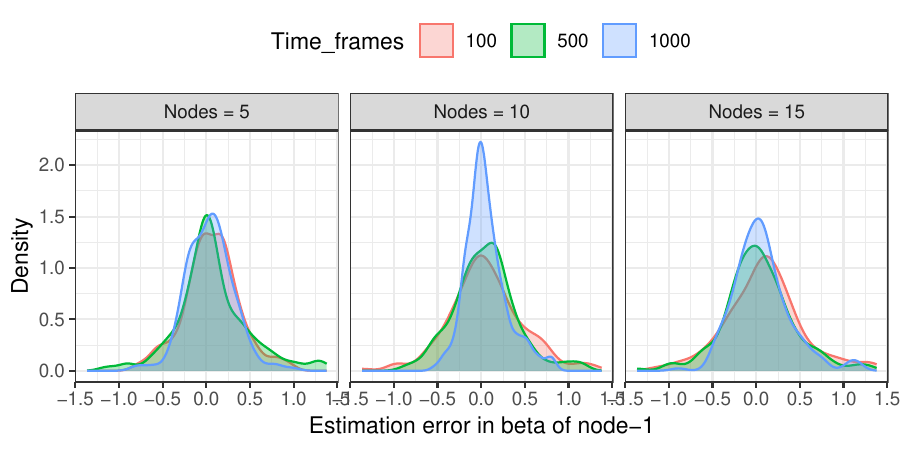}
    \caption{Static model. Estimation errors across 100 different network datasets: Left panel shows the error for the parameter $\alpha$, and the right panel shows the error for the parameter $\boldsymbol{\beta}$, with node counts $N =\{5, 10, 15\}$ and time points $T = \{100, 500, 1000\}$.}
    \label{fig: Density_plot_estimation_err}
\end{figure}

\begin{figure}[htbp]
    \centering    
    \includegraphics[width = 0.496 \textwidth]{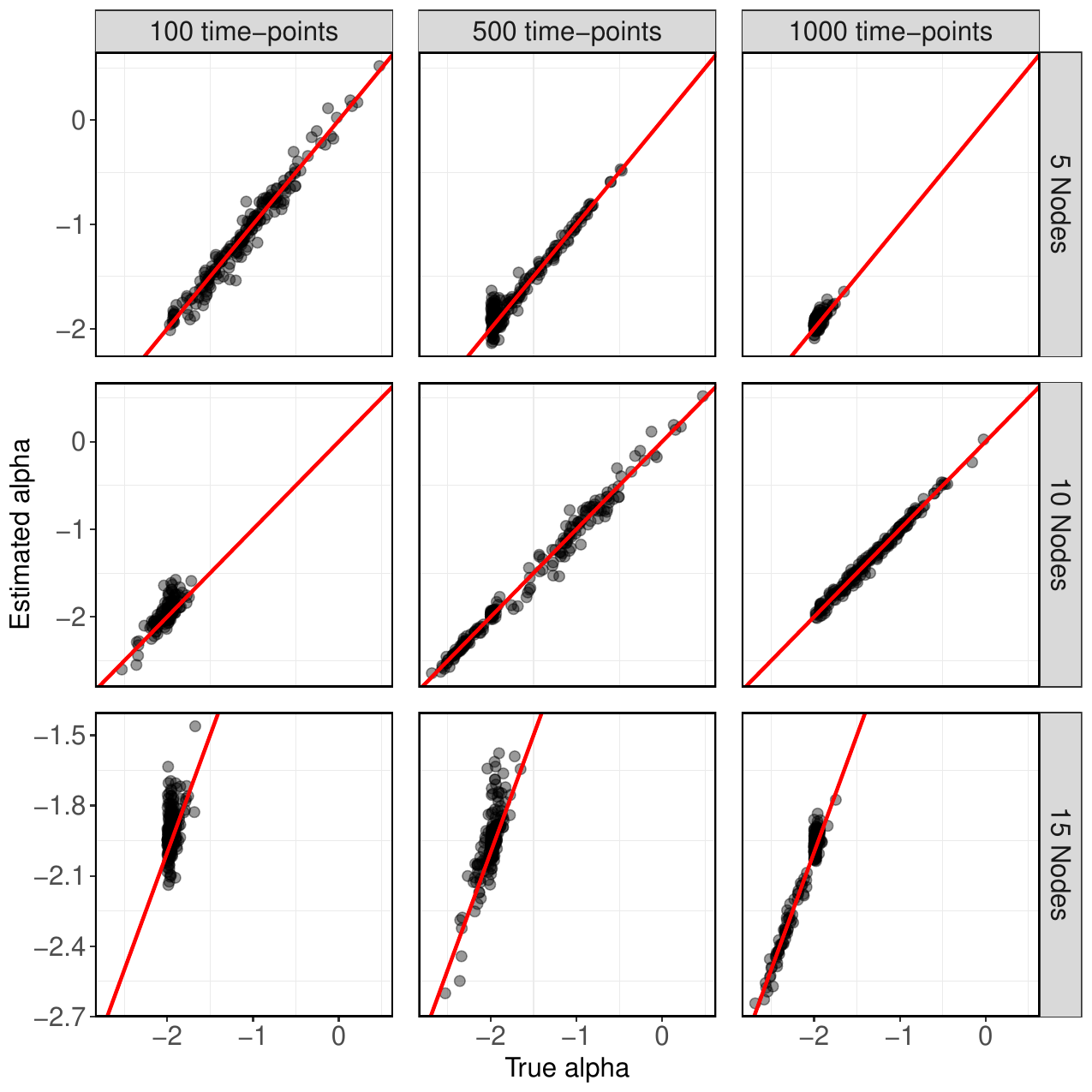}
    \includegraphics[width = 0.496 \textwidth]{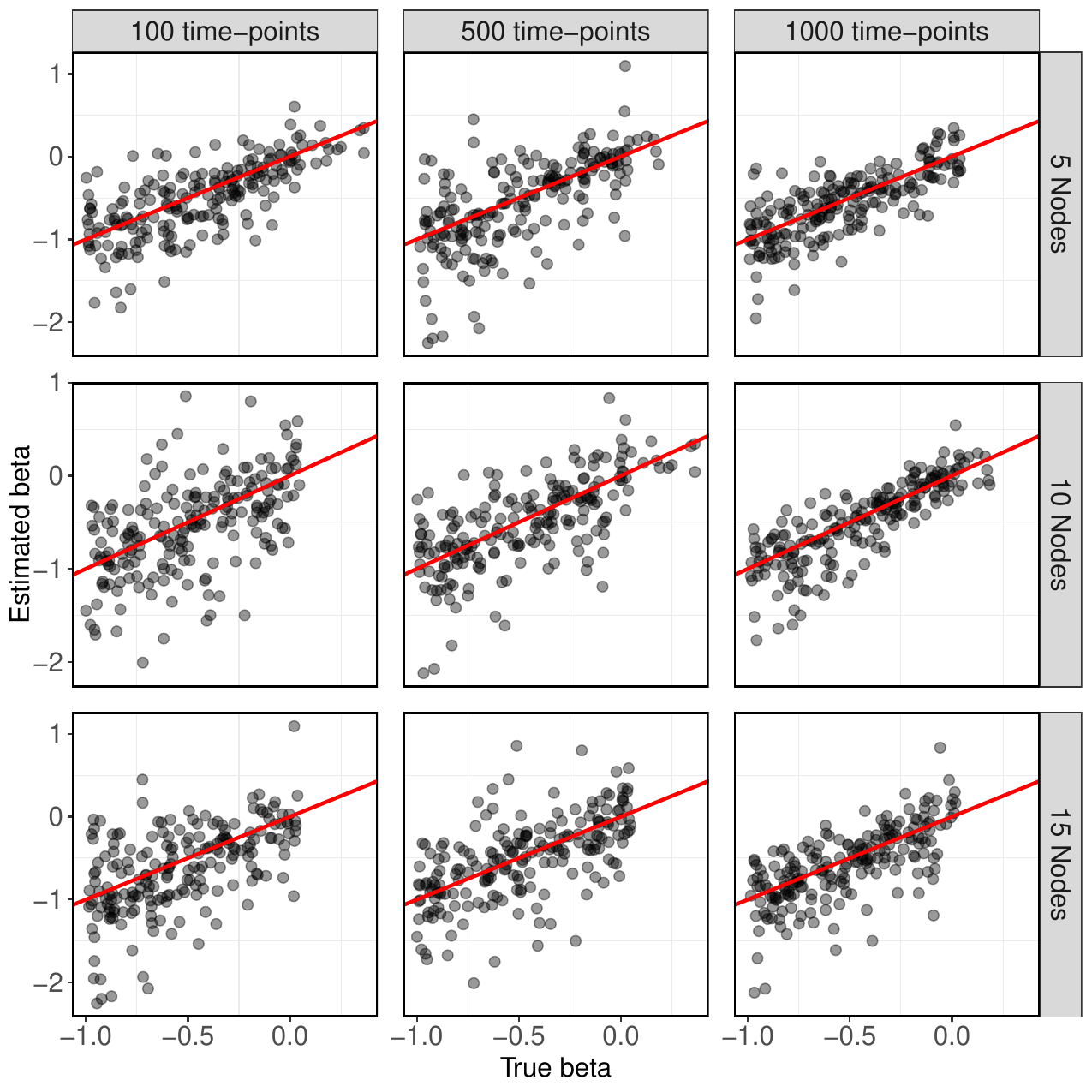}
    \caption{Static model. True values (x-axis) versus estimated values (y-axis) for the parameters across 100 different network datasets with nodes $N =\{5, 10, 15\}$ and time points $T = \{100, 500, 1000\}$. Left: $\alpha$, Right: $\beta$}

    \label{fig: Scatter_plot_true_vs_Est_val}
\end{figure}

We illustrate the estimation error for the parameters $\alpha$ and $\boldsymbol{\beta}$ using the density plots in Figure~\ref{fig: Density_plot_estimation_err}. 
Similarly, we report a comparison between true and estimated values in Figure~\ref{fig: Scatter_plot_true_vs_Est_val}.
The simulated results lead to several key conclusions. As shown in the density plots in Figure \ref{fig: Density_plot_estimation_err}, there are observable trends as $N$ and $T$ increase. Specifically, the estimation errors for both parameters $\alpha$ and $\boldsymbol{\beta}$ decrease as the number of time frames grows grows. Similar insights can be gathered from the scatter plots in Figure \ref{fig: Scatter_plot_true_vs_Est_val}, where smaller sample sizes result in greater dispersion of the estimated values, while larger sample sizes yield estimates that are closer to the true values. In all plots, we observe that the error decreases when the number of time frames increases. For $\alpha$, the error decreases also when the number of nodes increase, or when the number of time frames and number of nodes jointly increase. For $\beta$, we do not notice any particular improvement when the number of nodes increases. This is reasonable since each $\beta$ parameter is node-specific, and so adding more nodes also increases the number of parameters to estimate. Overall, the results demonstrate the effectiveness of the L-BFGS optimizer in providing reliable parameter estimates, even with relatively limited data. Our results are in agreement with the recent evidence also shown for a similar model in \textcite{Kaur_rastelli_2024}.

\begin{figure}[htbp]
    \centering    
    \includegraphics[width = 0.8 \textwidth]{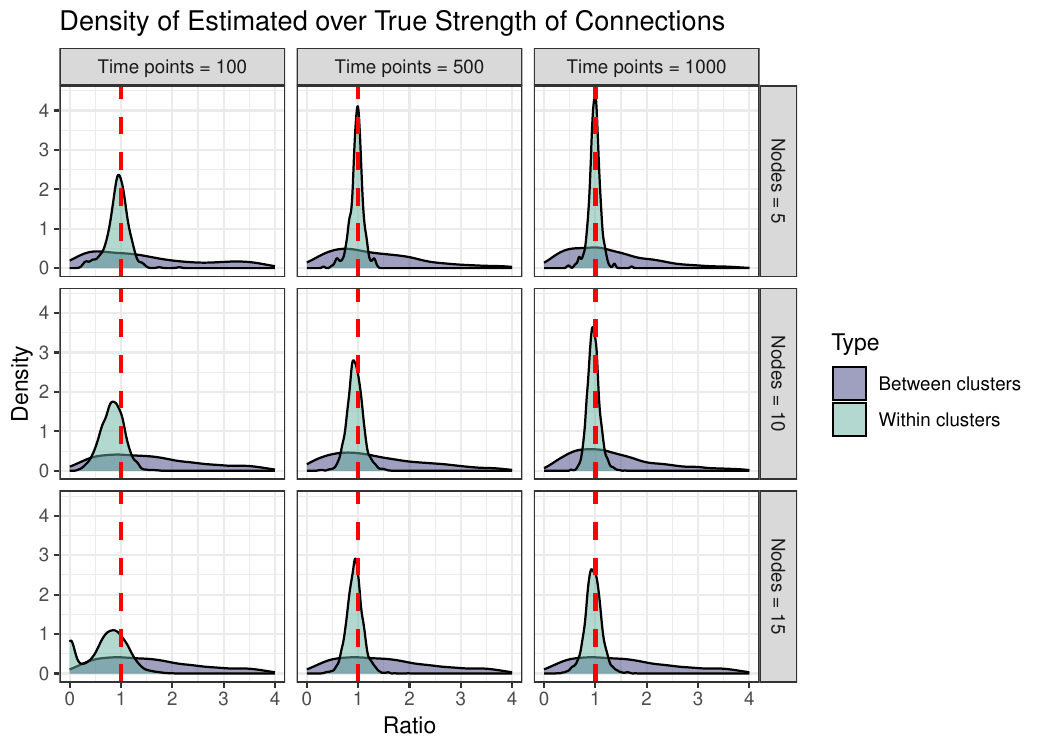}
    \caption{Distribution of pairwise interaction ratios between and within clusters, comparing estimated latent positions with true latent positions among $100$ different network datasets with nodes $N =\{5, 10, 15\}$ and time points $T = \{100, 500, 1000\}$. }
    \label{fig: density_plot_true_vs_Est_conn}
\end{figure}

Similarly to \textcite{Kaur_rastelli_2024}, we assess the estimation accuracy for the latent positions using the ratios between estimated distances and true distances. The plot in Figure \ref{fig: density_plot_true_vs_Est_conn} presents the distributions of the ratios separately for interactions that happen between-cluster and within-cluster.
This method of plotting ratios is also employed by \textcite{Sewell20151646, Sewell2016105} to compare the estimated latent space with the true latent space, bypassing any potential rotation or translation.
A narrow distribution plot centered around 1 suggests that the estimated latent space closely matches the true latent space, up to a scaling factor.

The distribution of ratios for within-cluster interactions is consistently centered around 1, which is indicated by the red dotted line in the plot. Additionally, as the sample size increases, these distributions become more tightly concentrated around 1. This pattern is consistent across different network sizes. Conversely, the distribution of ratios for between-cluster interactions appears relatively flat. These results emphasize a critical feature of the model, whereby dependency information is required to accurately estimate the latent positions; conversely, when series tend to be independent, the model struggles to recover the latent space because there is no information to be used.

\subsection{Dynamic time series distance model} \label{Dynamic_DTSLPM}

In this section, we apply our methodology to simulated data in order to assess the temporal performance of the DTSLPM model through a series of experiments. Our primary objective is to assess the model's capability in capturing network connections. It is important to note that the goal is not to recover the exact trajectories but rather to estimate trajectories that yield approximately similar interconnections between the nodes.

In this section, since we consider one individual dataset at a time, we proceed to apply a fully Bayesian procedure using HMC to obtain estimates of the latent node positions $\boldsymbol{\cal Z}_{t}$ and other model parameters. With regards to the random walk process prior on the latent trajectories, the initial and subsequent standard deviations of the increments are modeled using
\begin{gather*}
{\rho} \sim Gamma(2,2) \\
{\sigma} \sim Gamma(5,10) \\
\end{gather*}
similarly to \textcite{Sewell20151646, Sewell2016105, Friel20166629}.
For each dataset, we assess convergence through visual inspection of the trace plots, along with the examination of the $\hat{R}$ statistic, also known as the potential scale reduction factor \parencite{BDA_Gelman2013}. To ensure the quality and reliability of the parameter estimates, we also consider the Effective Posterior Sample Size (ESS).

\subsubsection{Experiment 1}
In this initial experiment, the data-generating mechanism is designed to capture the time-varying interrelationships between nodes within clusters. We note that, under a dynamic latent space setting, the definition of cluster becomes more blurred, since nodes can move along trajctories and thus change community or exhibit other types of dynamic behaviors. We consider a network of four nodes observed over 200 time points, and we let the nodes move in non-trivial ways, but nonetheless following clear regular patterns. 

\begin{figure}[H]
    \centering    
    \includegraphics[width = 0.82 \textwidth]{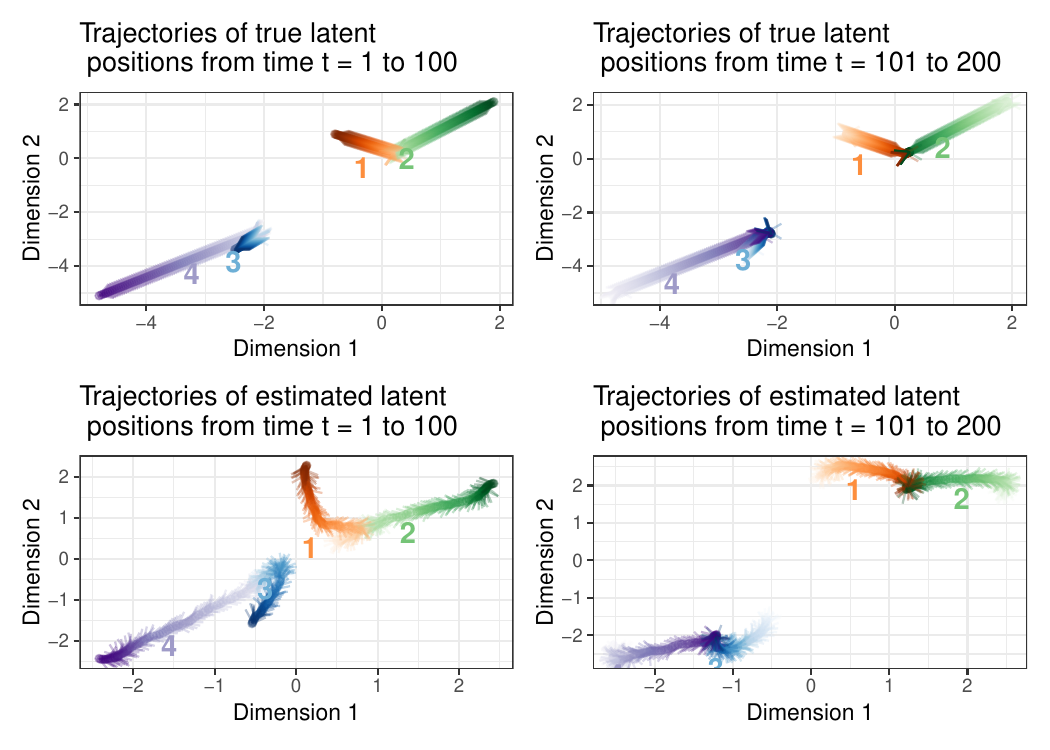}
    \caption{Posterior means of latent node positions, with each color representing a different node and arrows indicating the temporal evolutions of the trajectories. The left panel shows the trajectories of each node during the first half of the time period (t = 1 to 100), while the right panel displays the trajectories during the second half (t = 101 to 200).}
    \label{fig: EXP1 trajectory plot1}
\end{figure}

The data generation process is as follows:
\begin{gather*}
    \alpha \sim Uniform(-3,3) \notag \\
     \boldsymbol{\beta} \sim Uniform(-1,1) \notag \\
\end{gather*}
The initial positions of nodes are generated such that they are grouped up in two pairs, with nodes changing their relative positions while remaining within their cluster (see top panels of Figure \ref{fig: EXP1 trajectory plot1}). Thus, the objective of this experiment is to track node-specific activity within the clusters. 

In this experiment, we obtain a total of $5,000$ iterations, with a burn-in period of $3,000$ iterations and a thinning rate of $2$, resulting in $1,000$ samples to represent the posterior distributions of the model parameters. 
Table \ref{tab:Posterior_estimates_exp1} presents the posterior mean estimates for two parameter values alongside their true values. 
\begin{table}[H]
\centering
\begin{minipage}[b]{0.6\linewidth}
\centering
\begin{tabular}{lrrr}
  \toprule
Parameter & True  & Mean & 95\% credible interval \\ 
  \midrule
  $\alpha$ & -0.31 & -0.20 & (-0.33,-0.07) \\ 
 $\beta_1$ & -0.86 & -0.87 & (-1.20,-0.57) \\ 
  $\beta_2$ & -0.25  & -0.12 & (-0.33,0.07) \\ 
  $\beta_3$ & -0.84  & -0.97 & (-1.30,-0.65) \\ 
  $\beta_4$ & -0.74 & -0.64 & (-0.94,-0.35) \\
   \bottomrule
\end{tabular}
\end{minipage}
\caption {Posterior means with 95\% credible intervals for experiment 1.}
\label{tab:Posterior_estimates_exp1}
\end{table}
In Figure \ref{fig: EXP1 trajectory plot1}, we display two separate plots, each comparing the true and estimated node-specific cluster activities. 
The left plot corresponds to the first half of the time period, while the right plot represents the second half, showing the corresponding latent positions. These plots highlight the strong alignment between the estimated and true node-specific trajectories.

\begin{figure}[H]
    \centering    
    \includegraphics[width = 0.82 \textwidth]{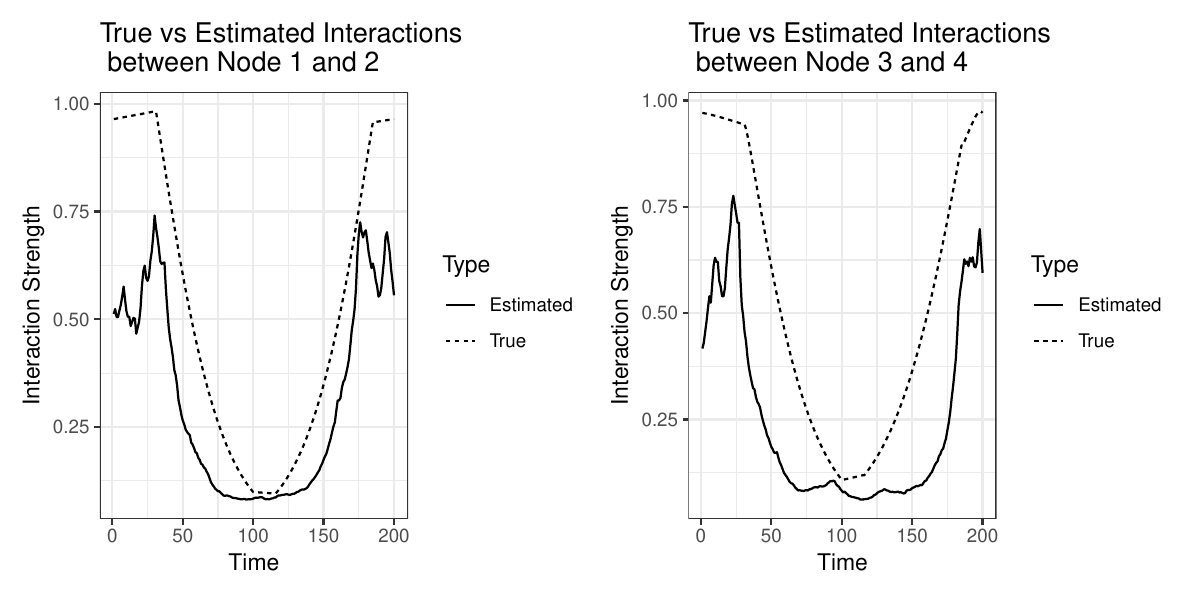}
    \caption{Plot of the posterior means of the estimated pairwise connections between nodes, shown as solid lines, and compared with the true pairwise connections, represented as dotted lines. The left panel compares the connection between nodes 1 and 2, while the right panel compares the connection between nodes 3 and 4.}
    \label{fig: exp1_interactions}
\end{figure}

Another key objective of this experiment is to capture the intensity of interconnections between nodes within the clusters. Figure \ref{fig: exp1_interactions} demonstrates that the estimated interactions closely follow the true interactions. 

\begin{figure}[H]
     \centering    
     \includegraphics[width = 0.9 \textwidth]{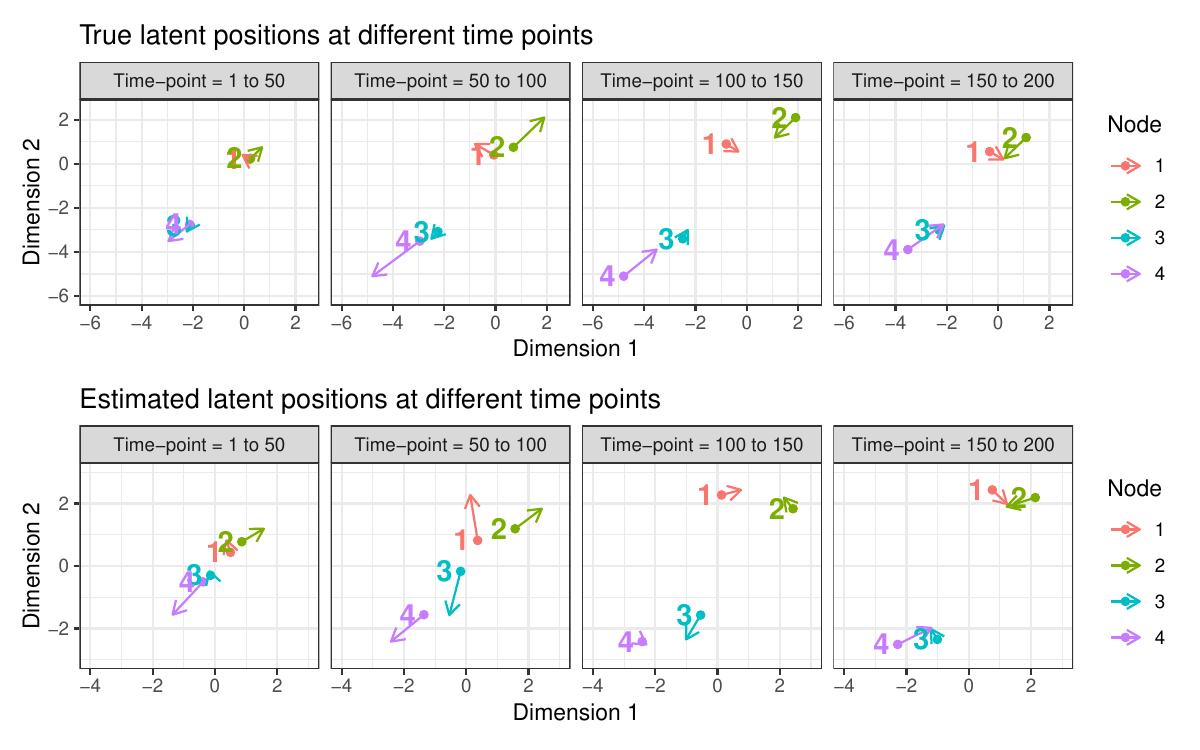}
     \caption{Posterior means of true and estimated latent node positions at different time points, with each color representing a different node and arrows indicating the temporal evolutions of the trajectories.}

     \label{fig: Movement_at_tps_exp1_true_est}
 \end{figure}

Snapshots of the true and estimated latent positions are also provided in Figure \ref{fig: Movement_at_tps_exp1_true_est}, where we see a good agreement thus indicating accurate estimation.

\subsubsection{Experiment 2}
In this second experiment, we utilize a data-generating method that is designed to measure the temporal interactions of a node as it transitions between clusters. The experiment involves a network of four nodes observed over 200 time points. The data-generating process for this experiment is identical to that used in Experiment 1, however we create different trajectories in this case. The primary objective is to track the temporal interactions and trajectory of node 1 as it transitions between two clusters.

\begin{figure}[H]
    \centering    
    \includegraphics[width = 0.8 \textwidth]{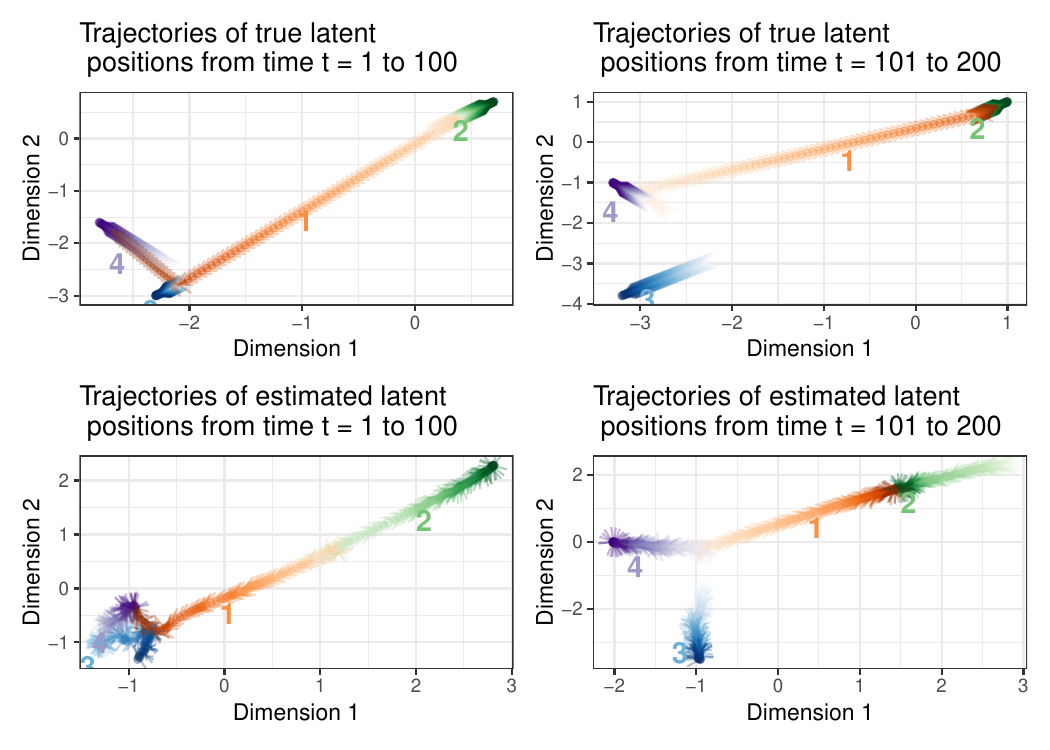}
    \caption{ Posterior means of latent node positions, with each color representing a different node and arrows indicating the temporal evolutions of the trajectories. The left panel shows the trajectories of each node during the first half of the time period (t = 1 to 100), while the right panel displays the trajectories during the second half (t = 101 to 200).}
    \label{fig: EXP2 trajectory plot1}
\end{figure}

\begin{figure}[H]
\centering    
\includegraphics[width = 0.9 \textwidth]{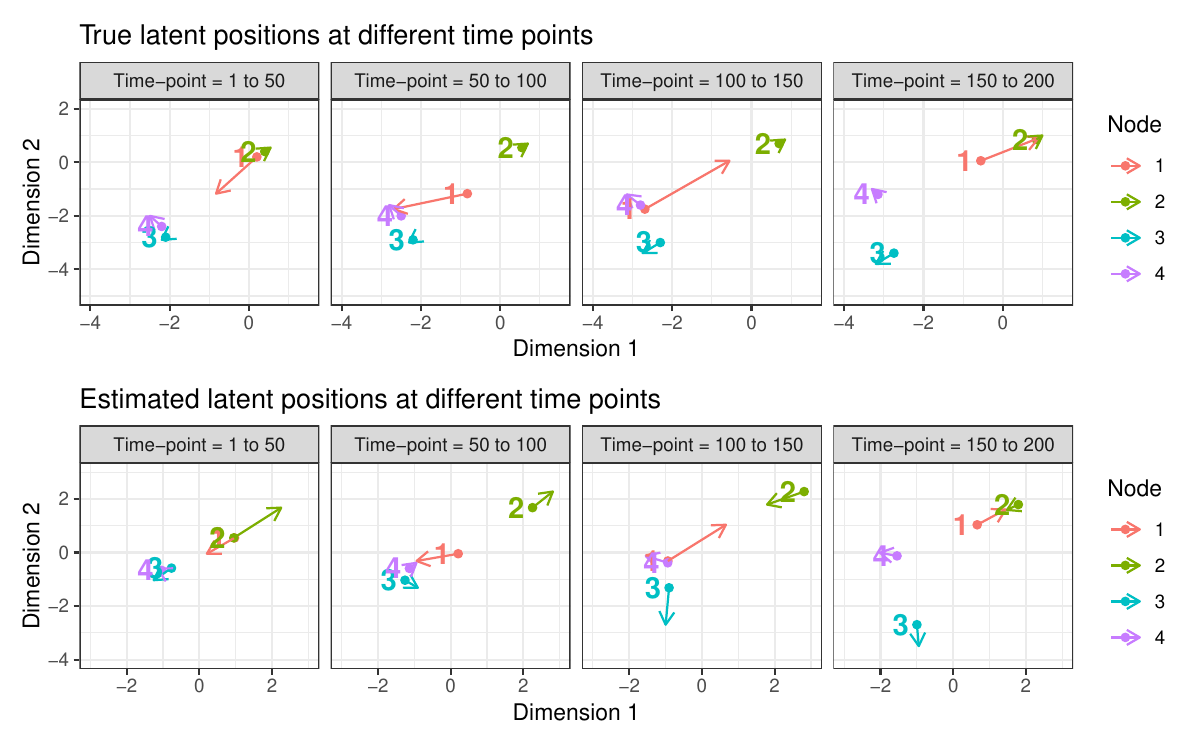}
\caption{Posterior means of true and estimated latent node positions at different time points, with each color representing a different node and arrows indicating the temporal evolutions of the trajectories.}

\label{fig: Movement_at_tps_exp2_true_est}
\end{figure}

The HMC chains were run for $10,000$ iterations, with a burn-in period of $5,000$ iterations and a thinning rate of $5$, resulting in $1,000$ samples to represent the posterior distributions of the model parameters. 
The results for the posterior mean estimates of two parameter values, as well as the true parameter estimates, are presented in Table \ref{tab:Posterior_estimates_exp2}. 
\begin{table}[htbp]
\centering
\begin{minipage}[b]{0.6\linewidth}
\centering
\begin{tabular}{lrrr}
\toprule
Parameter & True  & Mean & 95\% credible interval \\ 
\midrule
$\alpha$ & -0.005 & 0.16 & (0.04,0.28) \\  
$\beta_1$ & -0.53   & -0.67 & (-0.79,-0.54) \\
$\beta_2$ & 0.01  & -0.04 & (-0.14,0.06) \\ 
$\beta_3$ & -0.67  & -0.74 & (-0.92,-0.57) \\ 
$\beta_4$ & -0.15 & -0.33 & (-0.46,-0.19) \\ 
\bottomrule
\end{tabular}
\end{minipage}
\caption {Posterior means with 95\% credible intervals for experiment 2.}
\label{tab:Posterior_estimates_exp2}
\end{table}
Figure \ref{fig: EXP2 trajectory plot1} displays two distinct plots, respectively displaying the actual and the estimated trajectories particular to each node. 
The fitted model demonstrates a robust ability to capture the trajectories for all nodes. The results, as depicted in Figure \ref{fig: EXP2 trajectory plot1}, highlight that the trajectory of node 1, which is of particular interest, aligns closely with the true trajectory. Figure \ref{fig: Movement_at_tps_exp2_true_est} further illustrates that the estimated movement of node 1 is well aligned with the its true movement.

\begin{figure}[H]
    \centering    
    \includegraphics[width = 0.9 \textwidth]{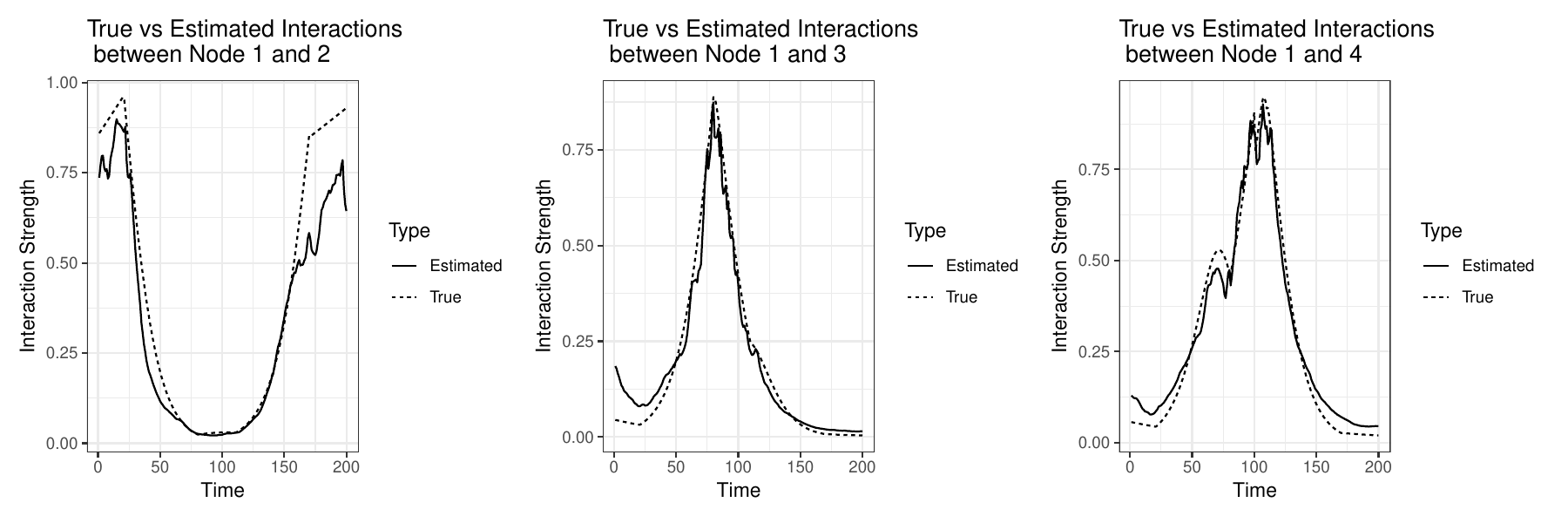}
    \caption{Plot of the posterior means of the estimated pairwise connections between nodes, shown as solid lines, compared with the true pairwise connections, represented as dotted lines. The left plot compares the connection between nodes 1 and 2, the middle plot compares the connection between nodes 1 and 3, and the right panel compares the connection between nodes 1 and 4. }
    \label{fig: exp2_interactions}
\end{figure}

The trajectories of the other nodes are also well recovered by the model. Indeed, the model performs well in recovering the interaction intensities between node 1 and the other nodes, as shows in Figure \ref{fig: exp2_interactions}.

\subsubsection{Experiment 3}

In the final experiment, the data-generating mechanism is intended to capture the temporal interactions between nodes as they simultaneously switch clusters. We analyze dynamic networks consisting of 4 nodes observed over 200 equally spaced time points. The data generation process in this experiment is the same as that used in Experiment 1, albeit with different trajectories. Initially, the nodes are divided into two groups, with nodes 1 and 4 then exchanging their positions. The main objective is to track how the connections between the nodes change as a result of these shifts in cluster membership.

The HMC algorithm was run for $50,000$ iterations, with the first $30,000$ iterations discarded as burn-in. A thinning rate of $20$ was applied, resulting in $1,000$ samples used to approximate the posterior distributions of the model parameters. 
Table \ref{tab:Posterior_estimates_exp3} shows strong correspondence in the true parameter estimates and the posterior mean estimates of model parameters $\alpha$ and $\boldsymbol{\beta}$. 
\begin{table}[htbp]
\centering
\begin{minipage}[b]{0.6\linewidth}
\centering
\begin{tabular}{lrrr}
  \toprule
Parameter & True  & Mean & 95\% credible interval \\ 
  \midrule
  $\alpha$ & 0.22 & 0.37 & (0.17,0.56) \\  
 $\beta_1$ & -0.67   & -0.70 & (-0.81,-0.58) \\
  $\beta_2$ & -0.15  & -0.19 & (-0.33,0.07) \\ 
  $\beta_3$ & -0.85  & -1.10 & (-1.30,-0.90) \\ 
  $\beta_4$ & -0.04 & -0.09 & (-0.16,-0.01) \\ 
   \bottomrule
\end{tabular}
\end{minipage}
\caption {Posterior means with 95\% credible intervals for experiment 3.}
\label{tab:Posterior_estimates_exp3}
\end{table}

\begin{figure}[htbp]
    \centering    
    \includegraphics[width = 0.82 \textwidth]{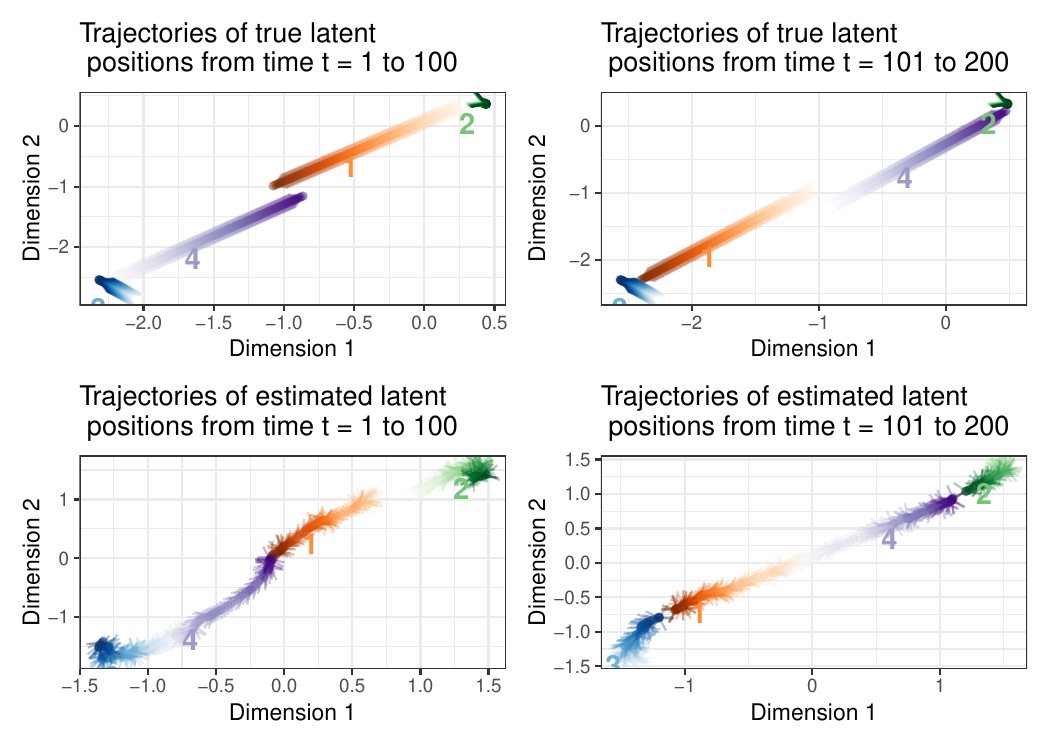}
    \caption{Posterior means of latent node positions, with each color representing a different node and arrows indicating the temporal evolution of the trajectories. The left panel shows the trajectories of each node during the first half of the time period (t = 1 to 100), while the right panel displays the trajectories during the second half (t = 101 to 200). }
    \label{fig: EXP3 trajectory plot1}
\end{figure}

\begin{figure}[htbp]
    \centering    
    \includegraphics[width = 0.82 \textwidth]{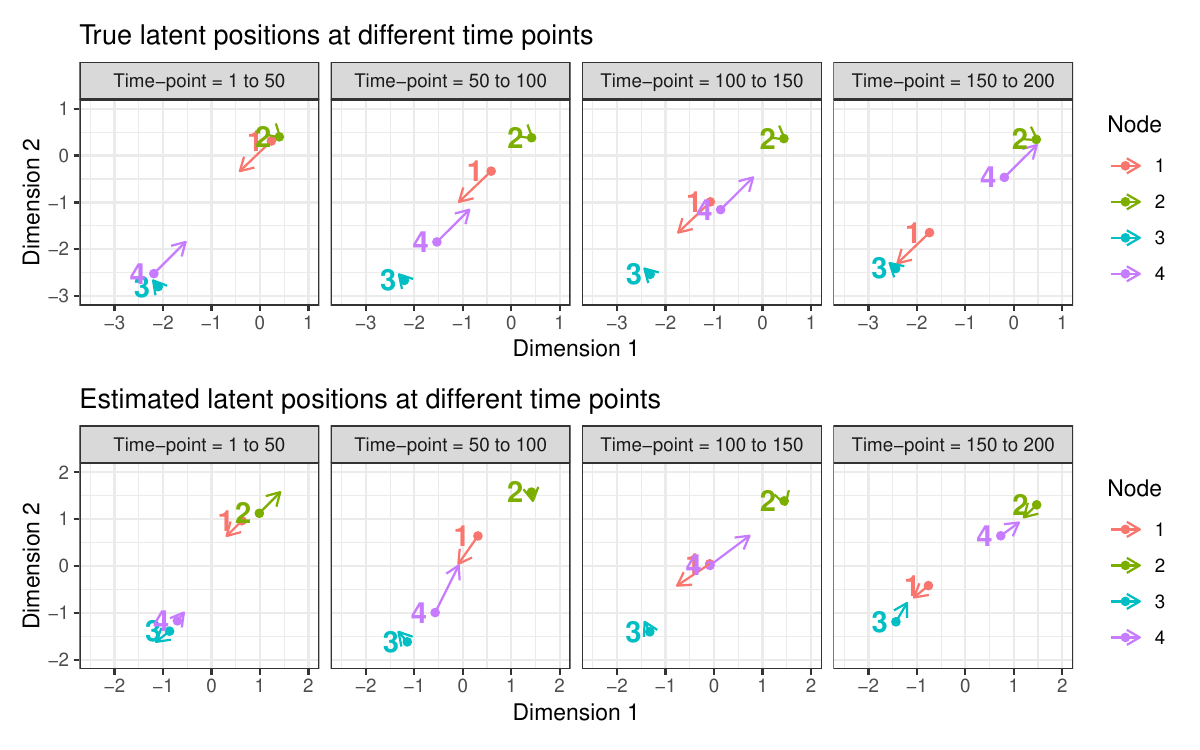}
    \caption{ Posterior means of true and estimated latent node positions at different time points, with each color representing a different node and arrows indicating the temporal evolutions of the trajectories.}
    \label{fig: Movement_at_tps_exp3_true_est}
\end{figure}

Figure \ref{fig: EXP3 trajectory plot1} presents the actual and estimated trajectories for each node. The fitted model shows a high level of accuracy in capturing the trajectories of nodes as they change clusters, as illustrated in the figure. Figure \ref{fig: Movement_at_tps_exp3_true_est}, which depicts the latent trajectories of each node at specific time points, further confirms that the estimated trajectories closely reflect the true trajectories, reinforcing the model's accuracy in tracking the movement of nodes over time.
 
\begin{figure}[htbp]
    \centering    
    \includegraphics[width = 0.82 \textwidth]{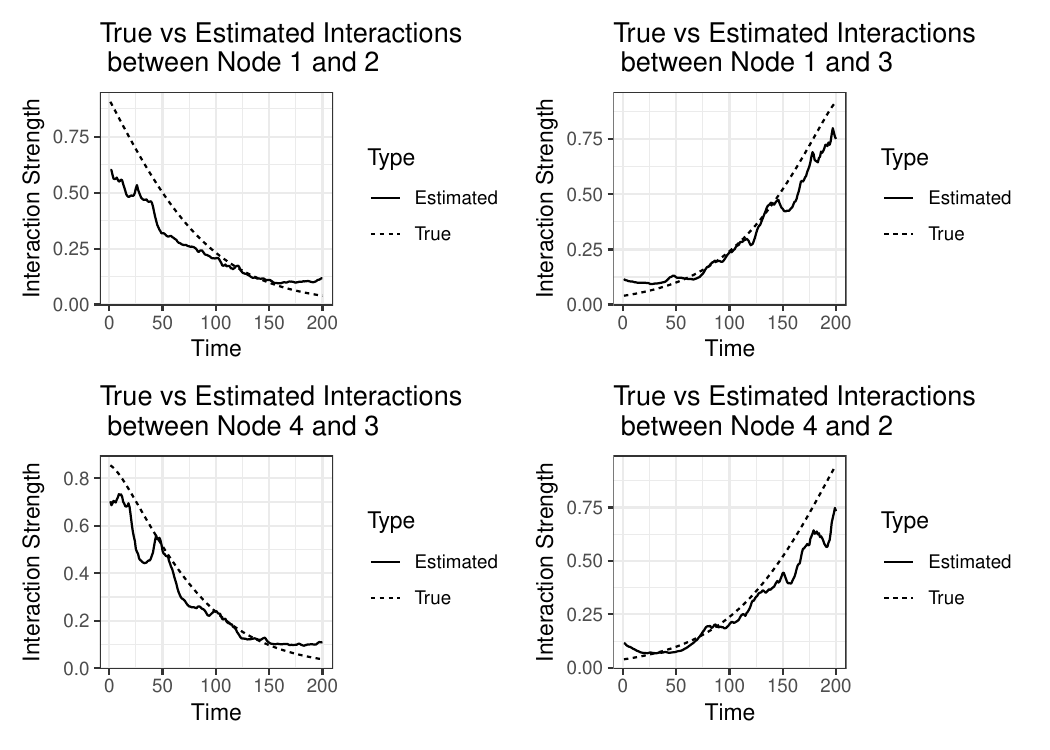}
    \caption{Plot of the posterior means of the estimated pairwise connections between nodes, shown as solid lines, and compared with the true pairwise connections, represented as dotted lines. The top panels compares the connections of node 1 with nodes 2 (left) and 3 (right), while the bottom panel compares the connections of node 4 with nodes 2 (left) and 3 (right).}
    \label{fig: exp3_interactions}
\end{figure}

In addition, the model performs well in recovering the connections between nodes with shifting cluster memberships. Figure \ref{fig: exp3_interactions} demonstrates that the estimated interconnections closely align with the true interconnections.

\section{Analysis of mumps cases} \label{real_dataset}

We study the number of cases of mumps in $N = 9$ regions of England recorded quarterly from $2008(Q1)-2020(Q2)$, for a total of 50 quarters. The data were collected from the Public Health England website\footnote{\url{https://www.gov.uk/government/organisations/public-health-england}}. Cases include those confirmed by oral fluid IgM antibody tests and by routine laboratory reports. 

\begin{figure}[htbp]
    \centering    
    \includegraphics[width = 0.72 \textwidth]{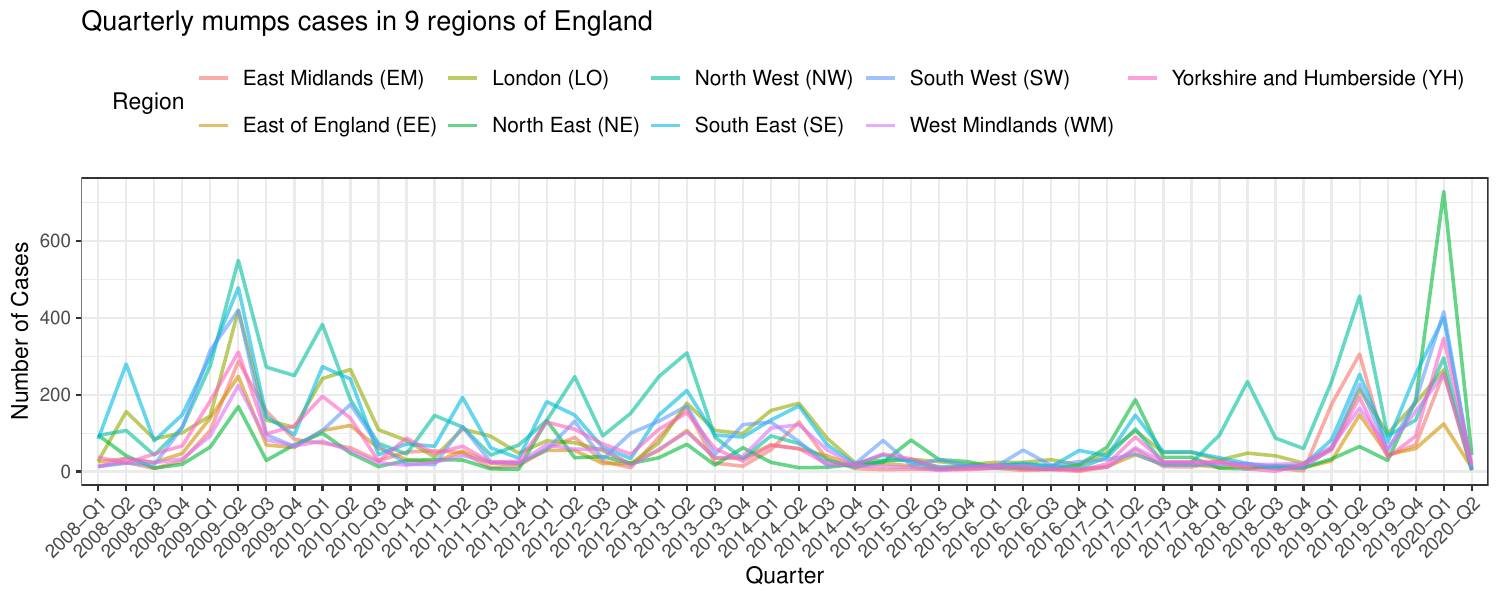}
    \includegraphics[width = 0.27 \textwidth]{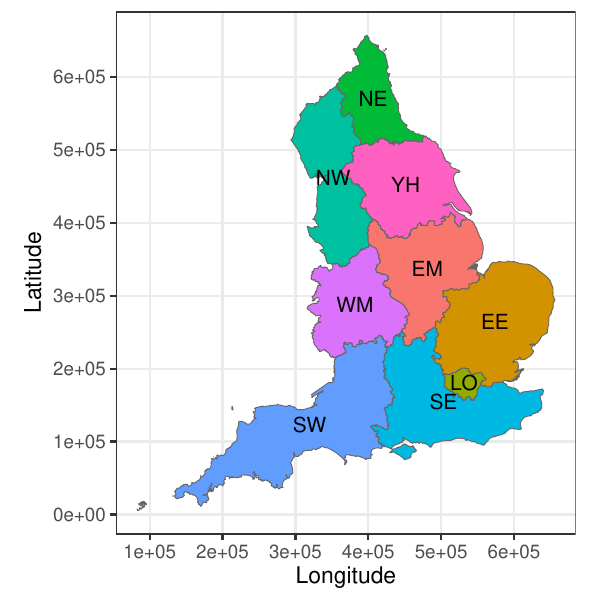}
    \caption{Left: quarterly number of mumps cases in 9 different regions of England. Right: regional breakdown of England.}
    \label{fig: ts_mumps_cases_england}
\end{figure}

\begin{figure}[htbp]
    \centering    
    \includegraphics[width = 1 \textwidth]{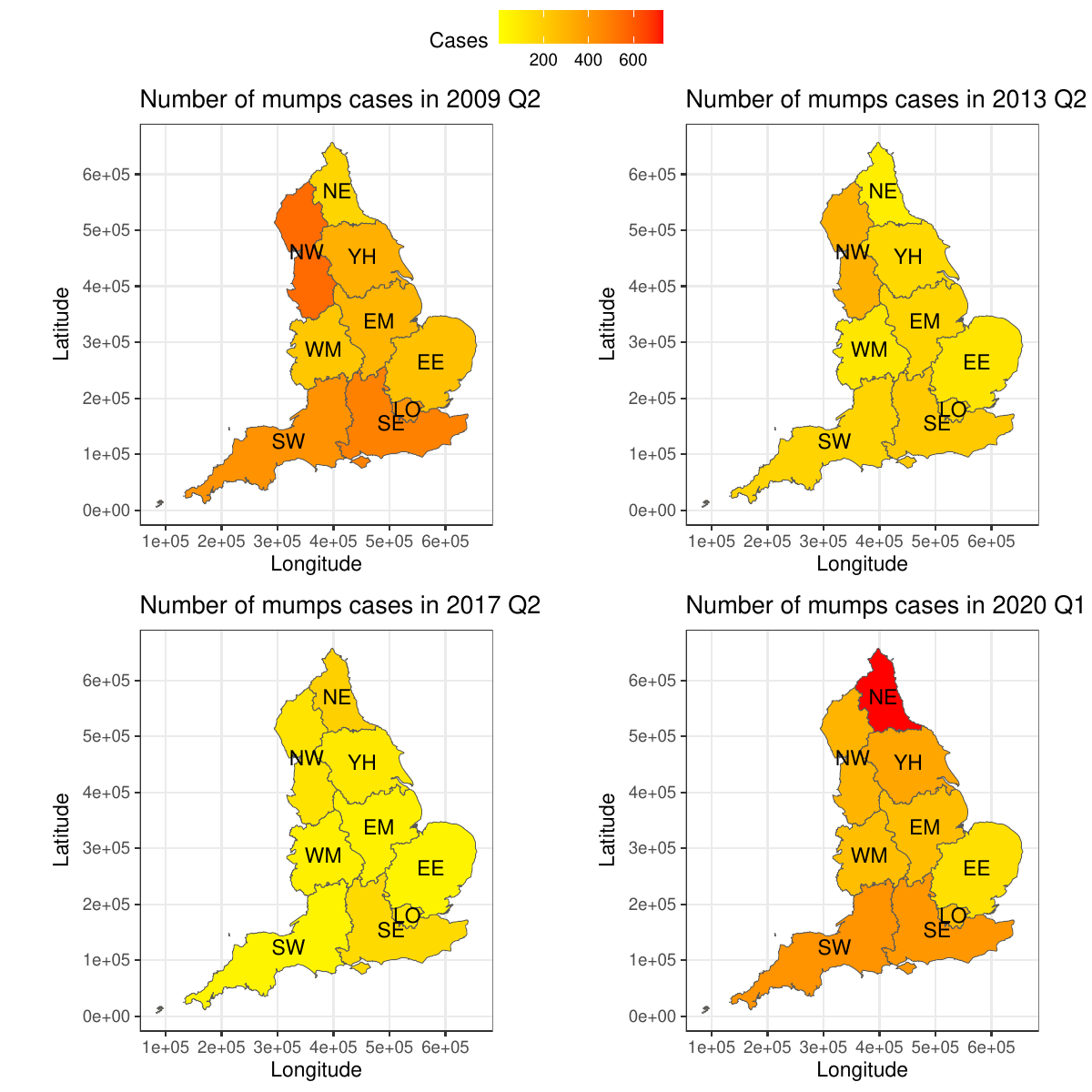}
    \caption{Distribution of mumps cases across regions of England: 2009 Q2 (top-left), 2013 Q2 (top-right), 2017 Q2 (bottom-left), and 2020 Q1 (bottom-right).}
    \label{fig: quarterly_mumps_regional_plot}
\end{figure}

These observed data can be written as $y_{i}^{t}$ for regions $i = 1, \dots, 9 $ over quarters $t = 1, \dots 50$: an illustration of it is provided in the left panel of Figure \ref{fig: ts_mumps_cases_england}. 
In the right panel, we present a geographical representation of England and its regions. Snapshots for 4 selected time points are shown in Figure \ref{fig: quarterly_mumps_regional_plot}, highlighting the periods with the highest peaks of mumps cases, with the number of cases being color-coded: deep red representing a high number of cases and yellow indicating fewer cases.
The goal of this analysis is to employ the DTSLPM model to highlight patterns of interactions between the number of mumps cases in different regions, at different times. These late    nt patterns can provide critical insights on possible contagion routes, and, ultimately, on the risk of an escalation of the spread of disease. 

For completeness, we fit both the static and the dynamic version of our model. We note that, in either model, we do not make use of any covariates, or geographical information on the regions. We point out that this additional information may definitely be used in our model, similarly to the work of \textcite{Kaur_rastelli_2024}. However, we do not consider this additional information within the scope of this work based on a latent variable framework. Indeed, the role of the latent positions that we estimate is to capture all hidden patterns, caused by geographical factors or otherwise. Additionally, we also note that the geographical information per se, is not necessarily the driving factor for contagion. Indeed, human mobility and human interactions may not necessarily be determined purely by the geographical distance, but rather, at least intuitively, they may be more related to transportation networks and other socio-economic factors that are difficult to observe and measure.

\subsection{Static model for the mumps cases}

We proceed by fitting the DTSLPM model as discussed in Section- \ref{static_DTSLPM}, to analyze the underlying static network that represents the interactions between regions.
We use HMC for fitting the DTSLPM and thus to obtain the posterior samples. A total of $30,000$ iterations were performed, with a burn-in period of $20,000$ iterations and a thinning rate of $10$, resulting in $1,000$ samples to represent the posterior distributions of the model parameters. 
For the model parameters $\alpha,  \boldsymbol{\beta}$, and $\boldsymbol{\cal Z}$, we employ non-informative priors, specifying independent normal priors with a mean of zero and a standard deviation of 10 for each. 

We use the $\hat{R}$ statistic and visual examination of the trace plots to check for convergence (\cite{BDA_Gelman2013}). To assess the quality and reliability of the samples of parameter estimates, the effective sample sizes were checked.
The results demonstrate good convergence, and the posterior summaries for all model parameters, except the latent positions, are presented in Table \ref{tab:mean_est_static_model}. 
\begin{table}[ht]
\centering
\caption{Posterior means with 95\% credible interval for the fitted static model.}
\label{tab:mean_est_static_model}
\begin{tabular}{lrrr}
  \toprule
Parameter & mean & 95\% credible interval \\ 
  \midrule
$\alpha$ & 2.18 & (2.12,2.23) \\ 
  $\beta_{EM}$ & 0.27 & (0.22, 0.33) \\ 
  $\beta_{EE}$ & 0.32 & (0.16,0.46) \\ 
  $\beta_{LO}$ & 0.34 & (0.28,0.41) \\ 
  $\beta_{NE}$ & 0.47 & (0.43,0.51) \\ 
  $\beta_{NW}$ & 0.51 & (0.48,0.56) \\ 
  $\beta_{SE}$ & 0.40 & (0.33,0.46) \\ 
  $\beta_{WM}$ & 0.32 & (0.25,0.39) \\ 
  $\beta_{YH}$ & 0.18 & (0.13,0.24) \\ 
  $\beta_{SW}$ & 0.29 & (0.24,0.33) \\ 
   \bottomrule
\end{tabular}
\end{table}
We observe that all the $\boldsymbol{\beta}$ parameters are positive, indicating positive autocorrelations. The North West (NW) region shows the strongest persistence in mumps cases over time within the same region, while Yorkshire and Humberside (YH) exhibit the weakest persistence.

We now turn to a detailed analysis of the pairwise relationships between the regions, which constitutes a key aspect of this study. The inclusion of the interaction term $\gamma_{ij}$ in our model enables us to investigate these regional interactions, using the spatial positioning $\boldsymbol{\cal Z}$ as a central element for interpreting these connections. This approach provides both quantitative data and visual representations, allowing us to identify specific patterns in the spread of mumps. 

The HMC output resulted in $1000$ posterior samples of latent positions $\boldsymbol{\cal Z}$ for each region of England. To post-process these latent positions for analysis, we applied the Procrustes transformations as discussed in Section \ref{Procrustes}, using the MAP as a reference for orientation.  

\begin{figure}[htbp]
    \centering    
    \includegraphics[width = 0.45 \textwidth]{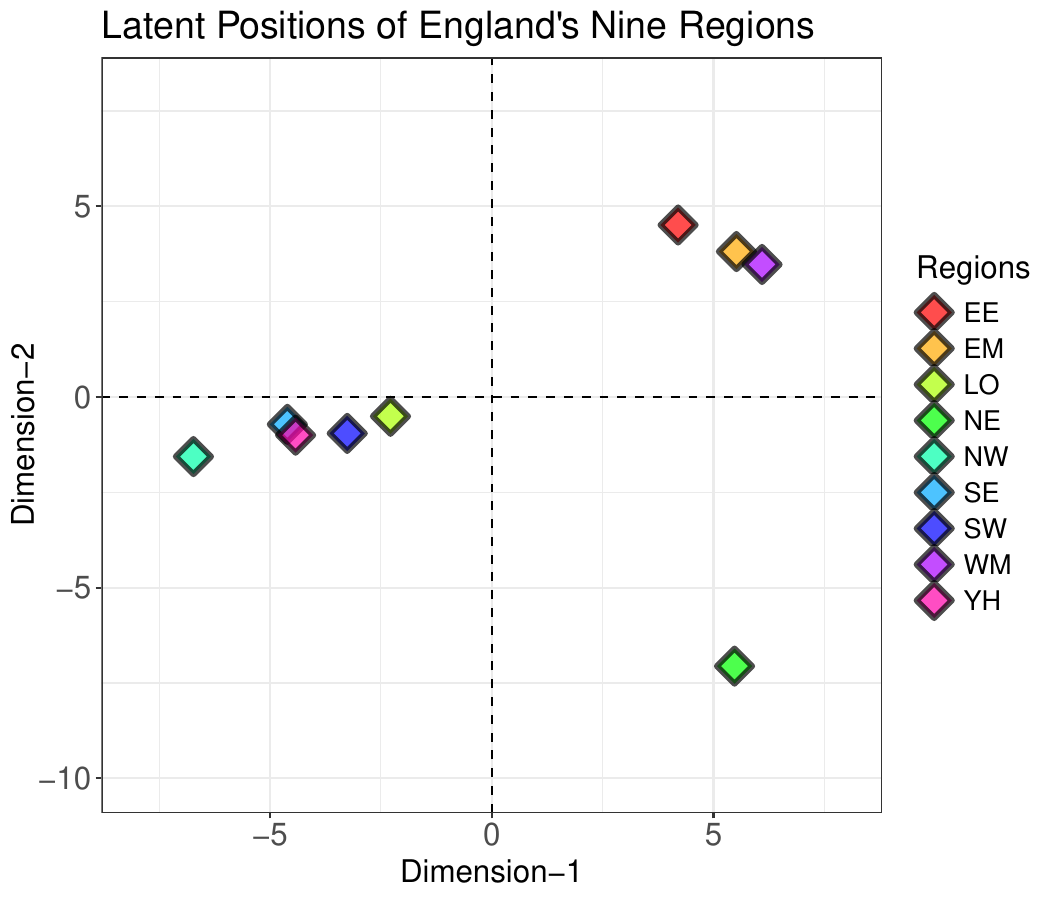}
    \includegraphics[width = 0.45 \textwidth]{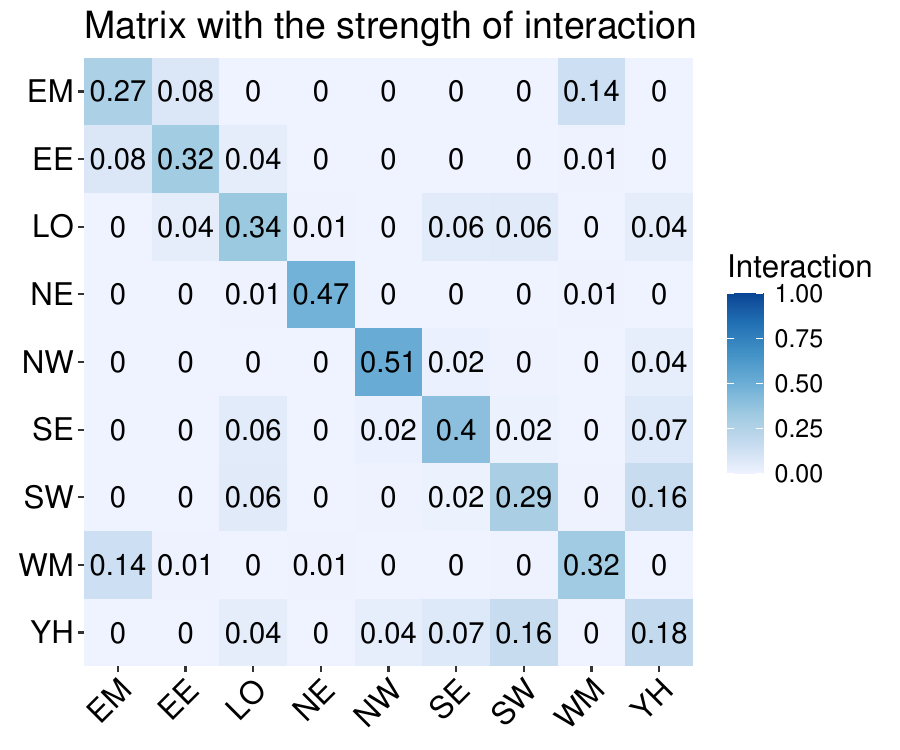}
    \caption{Left: latent space representation showing the posterior mean of locations for each region in England. Right: A matrix of connections between the different regions of England.}
    \label{fig: static_lp_and_interactions}
\end{figure}

The left panel of Figure \ref{fig: static_lp_and_interactions} shows the estimated regional positions, derived by calculating the posterior means of the latent positions for each region, following the application of the Procrustes transformation. 
In the right panel, the posterior average of the interactions matrix is presented, where the off-diagonal elements represent the pairwise interaction terms ($\boldsymbol{\gamma}$s), and the diagonal elements capture the autoregressive effects ($\boldsymbol{\beta}$s). Together, these plots provide insights into how the latent space models the interdependencies among the time series.

The right panel of Figure \ref{fig: static_lp_and_interactions} reveals distinct patterns, with the North West (NW), North East (NE), and South East (SE) exhibiting the strongest autoregressive effects, while London (LO), East of England (EE), East Midlands (EM), West Midlands (WM), and the South West (SW) exhibit moderate to weaker effects. Yorkshire and Humberside (YH) show the weakest autoregressive effect. These patterns may reflect the differing historical levels of mumps cases across regions.

The latent positions shown in the left panel of the same Figure reveal that Yorkshire and Humberside (YH), as well as the South West (SW), are positioned closely in the latent space, as are East Mainlands (EM) and West Mainland (WM). This proximity reflects the interconnection between these regions, also quantified by the non-diagonal elements on the right panel of Figure \ref{fig: static_lp_and_interactions}.
Although the overall connections are modest, the proximity of the nodes indicates subtle interactions in the spread of mumps between these regions, quantified on a scale from 1 (strong positive connection) to 0 (independence). The geographical proximity of East Mainlands (EM) and West Mainland (WM) corresponds to one of the strongest connections observed. Interestingly, despite being geographically distant, Yorkshire and Humberside (YH) and the South West (SW) show pronounced connections, too. These findings highlight spreading patterns that are not necessarily directly affected by geographical proximity.

\subsection{Results for dynamic time series distance model}

We extend now the analysis by fitting our full DTSLPM model in its dynamic version, in order to uncover the underlying temporal network and gain insights on mumps cases dynamics. 
The HMC algorithm was run for $300,000$ iterations, with a burn-in period of $250,000$ iterations and a thinning rate of $25$, resulting in $2,000$ posterior samples for the model parameters. Non-informative priors, as discussed in Section \ref{Prior_specification}, were used for the model parameters $\alpha$, $\boldsymbol{\beta}$, and $\boldsymbol{\cal Z}_{t}$. The remaining parameters related to latent positions are the hyperparameters $\rho$ and $\sigma$ for the movement of nodes that are user-specified. The hyperparameters are specified as follows: $\rho=10$ and $\sigma=0.05$.

The posterior summaries for all model parameters, except the latent positions, are presented in Table \ref{tab:mean_est_dynamic_model}. 
\begin{table}[ht]
\centering
\caption{Posterior means with 95\% credible intervals for the fitted dynamic model.}
\label{tab:mean_est_dynamic_model}
\begin{tabular}{lrrr}
  \toprule
Parameter & Mean & 95\% credible interval \\ 
  \midrule
  $\alpha$ & 1.45 & (1.02,1.83) \\ 
  $\beta_{EM}$ & -0.31 & (-0.49,-0.13) \\ 
  $\beta_{EE}$ & -0.60 & (-0.79,-0.17) \\ 
  $\beta_{LO}$ & -0.21 & (-0.34,-0.08) \\ 
  $\beta_{NE}$ & -0.12 & (-0.25,0.03) \\ 
  $\beta_{NW}$ & 0.02 & (-0.09,0.14) \\ 
  $\beta_{SE}$ & -0.03 & (-0.14,0.07) \\ 
  $\beta_{WM}$ & -0.46 & (-0.66,-0.12) \\ 
  $\beta_{YH}$ & -0.03 & (-0.16,0.10) \\ 
  $\beta_{SW}$ & -0.22 & (-0.34,-0.08) \\ 
   \bottomrule
\end{tabular}
\end{table}
We observe in this case that the majority of the $\boldsymbol{\beta}$ parameters are negative, indicating alternating positive and negative autocorrelations, which contribute to more unpredictable behavior. This result on the $\boldsymbol{\beta}$ parameters can be justified by the absence of any consistent upward or downward trend in mumps cases across the regions of England. Instead, the number of mumps cases in each region fluctuates, increasing and decreasing over time, which can be explained with a negative autoregressive coefficient. 

We now turn to a crucial aspect of the analysis: examining the pairwise relationships between the regions of England based on the estimated spatial positions $\boldsymbol{\cal Z}_{t}$, which evolve over time $t$. 
The model fit generated 2,000 posterior samples of the latent positions $\mathbf{z}_{it}$ for each $i^{th}$ region in England. For analysis, these latent positions were post-processed using Procrustes transformation, as outlined in Section \ref{Procrustes}, with the MAP estimate serving as the orientation reference. 

\begin{figure}[htpb]
    \centering    
    \includegraphics[width = 1 \textwidth]{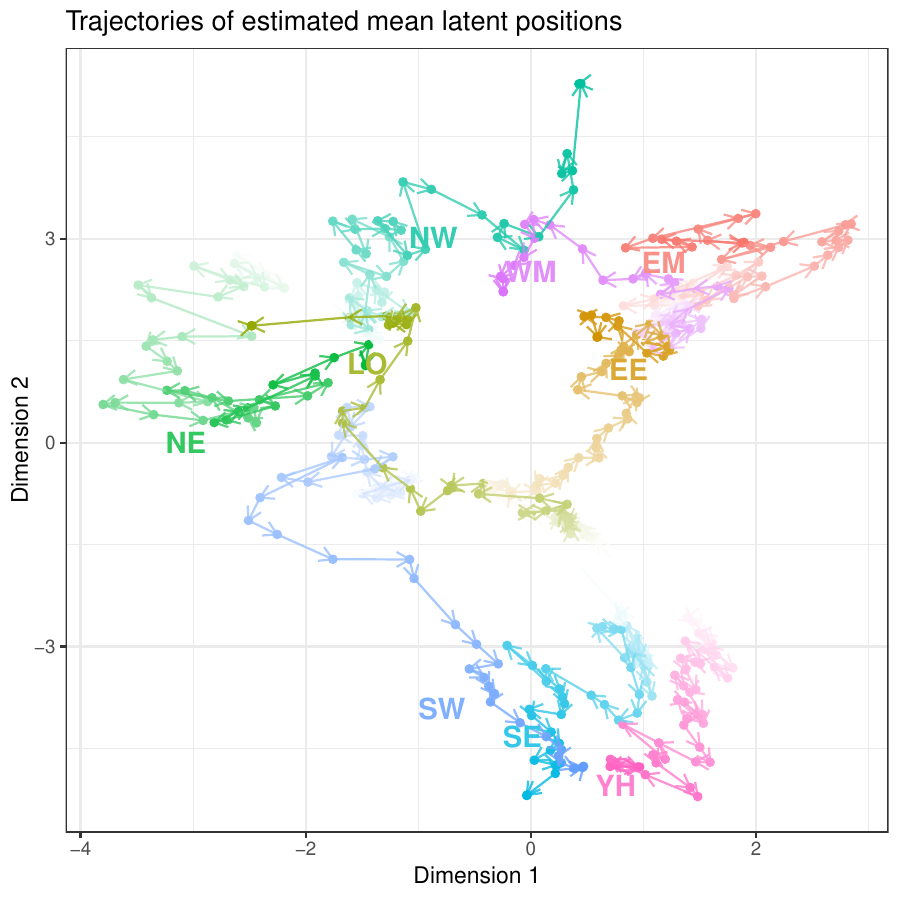}
    \caption{Posterior means of latent positions for each region in England. Each color represents a different region and arrows indicating the temporal evolutions of the trajectories.}
    \label{fig: dynamic_trajectory_mumps_cases}
\end{figure}

\begin{figure}[htbp]
    \centering    
    \includegraphics[width = 1 \textwidth]{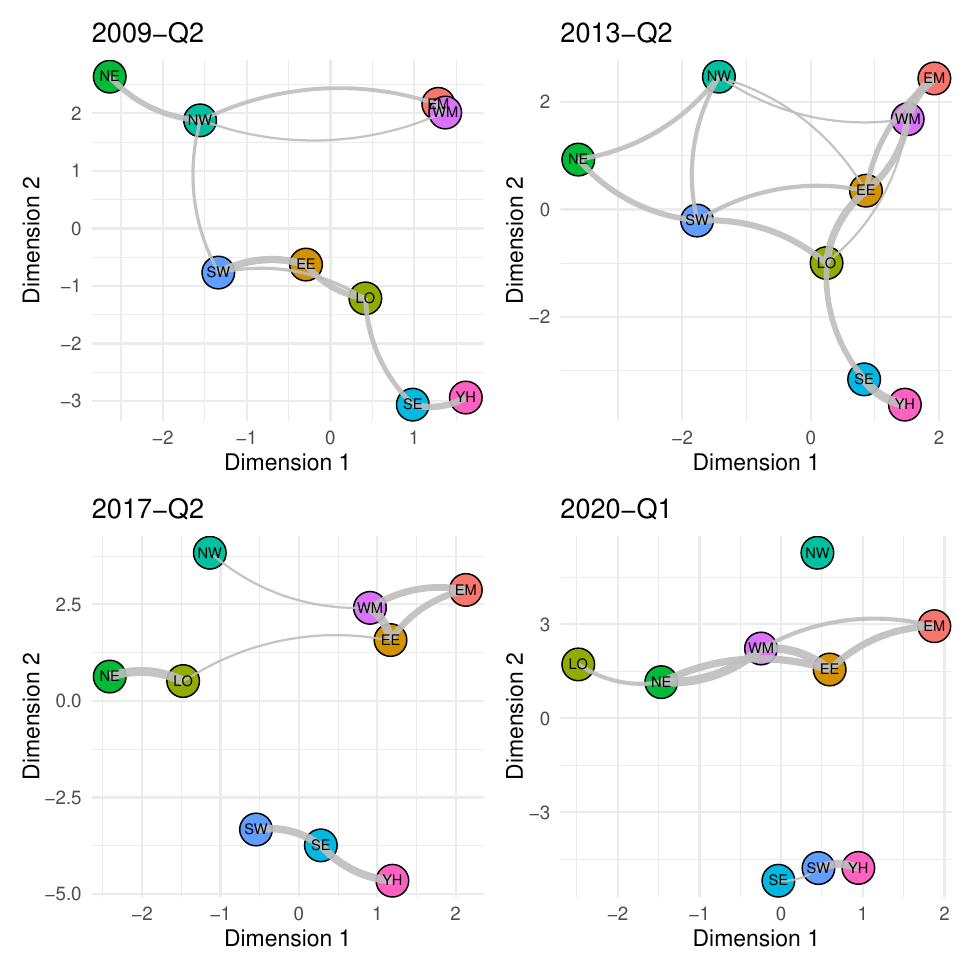}
    \caption{England's mumps cases time-stamped network: 4 networks representing the posterior mean of latent location of each region and size of the edge represents the strength of the connection between the regions. The colors of the nodes represent different regions.}
    \label{fig: Latent space based connections in region}
\end{figure}

\begin{figure}[htbp]
    \centering 
    \includegraphics[width = 1 \textwidth]{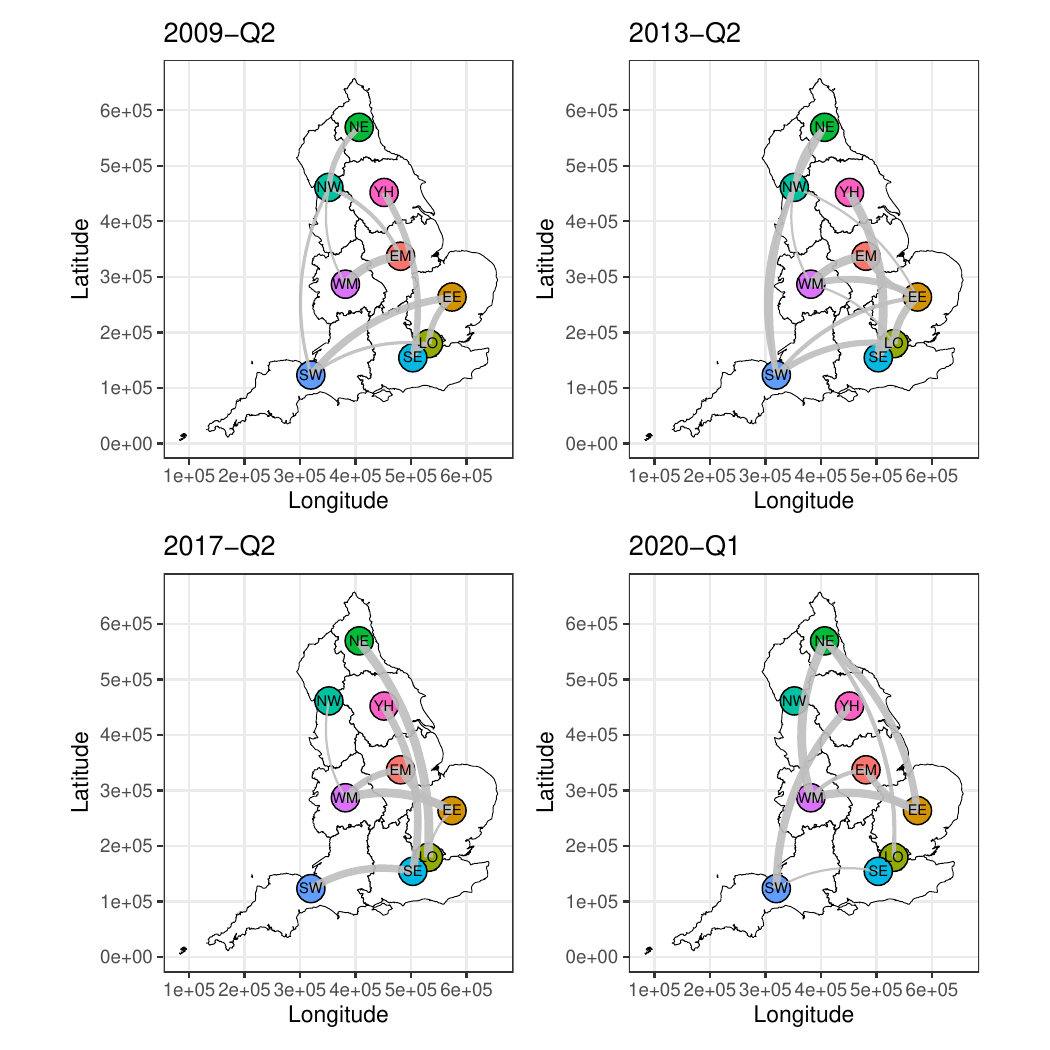}
    \caption{England's mumps cases time-stamped network: 4 networks display each region at its true geographic position, with edge thickness representing the strength of connections between regions. The colors of the nodes represent different regions.}
    \label{fig: geographical connections in region}
\end{figure}

Figure \ref{fig: dynamic_trajectory_mumps_cases} illustrates the posterior means of the latent locations of each region in England in each quarter from 2008 to 2020. 
The figure provides a visual representation of the hidden patterns of interconnections between the nodes and how they evolve with time based on the proximity of their locations in the latent space. The trajectories obtained by each region reveal some geographical patterns. 
We break down the information contained in this plot by focussing on four selected time points, and show the corresponding latent spaces in Figure \ref{fig: Latent space based connections in region}.
Similarly, and for completeness, the latent interactions are also show in Figure \ref{fig: geographical connections in region}, but using the actual geographical positions of the regions. 
Based on the data from 2008 to 2020, there is evidence of a strong connection in the spread of mumps infection between the East Midlands (EM) and West Midlands (WM) and their neighboring region in England. During the initial phase, the transmission dynamics between these regions were notably strong, indicating a significant level of infectious spread. However, a decline in direct transmission or spread between these regions is observed as time progresses. Additionally, a chain of transmission is observed, where mumps infections appear to be indirectly linked between West Midlands (WM) and East of England (EE) through the East Midlands (EM) region. Notably, the connections between East Midlands (EM) and East of England (EE), as well as West Midlands (WM) and East of England (EE), have been strengthening over time, highlighting the dynamic nature of mumps spread in these areas.

Despite not being geographically connected, there is a significant link in the spread of mumps cases between the South East (SE) region and Yorkshire and Humberside (YH) over the observed time period. This suggests indirect transmission pathways or shared risk factors driving the infection dynamics across these regions. Furthermore, from 2016 to 2020, Yorkshire and Humberside (YH), South West (SW), and South East (SW) formed a cluster of interconnected regions, with each influencing the spread of mumps in the others.

Before 2016, the South East (SE) region was notably affected by its neighboring region London (LO). However, the South West (SW) region exhibited significant connections with multiple regions, even those that were not its direct geographical neighbors. These connections evolved over time, linking the South West (SW) with East of England (EE), London (LO), North East (NE), and North West (NW). This pattern indicates a broader and dynamic network of mumps transmission, highlighting the role of indirect pathways in the spread of the infection across these diverse regions.
Additionally, the South East (SE) region has been notably affected by its neighboring areas, London (LO) and the South West (SW). The transmission dynamics indicate a strong epidemiological link with London (LO) until 2015, which then shifted towards a stronger connection with the South West (SW) in the subsequent years reflecting the shifting transmission routes.

An interesting pattern for the distribution of mumps cases across England is that the highest number of cases consistently occurs in the northernmost regions, specifically in the North East (NE) and North West (NW). Figure \ref{fig: quarterly_mumps_regional_plot} illustrates this trend, revealing that during the four peak periods depicted, the initial two periods exhibit the highest mumps cases in the North West (NW), while the last two peak periods show the highest cases concentrated in the North East (NE). This spatial distribution highlights the persistent vulnerability of these northern regions to mumps outbreaks over time.

Additionally, distinct patterns emerge in their connections with other regions during these peak periods. When the North West (NW) recorded the highest number of cases, it displayed significant connections with several regions, including North East (NE), East Midlands (EM), West Midlands (WM), and South West (SW), facilitating the spread of the infection over time. In contrast, during its peak periods, the North East (NE) primarily showed strong connections with the West Midlands (WM) and had a particularly notable link to London (LO), indicating focused transmission pathways between these areas.

\begin{figure}[hptb]
    \centering    
     \includegraphics[width = 0.8 \textwidth]{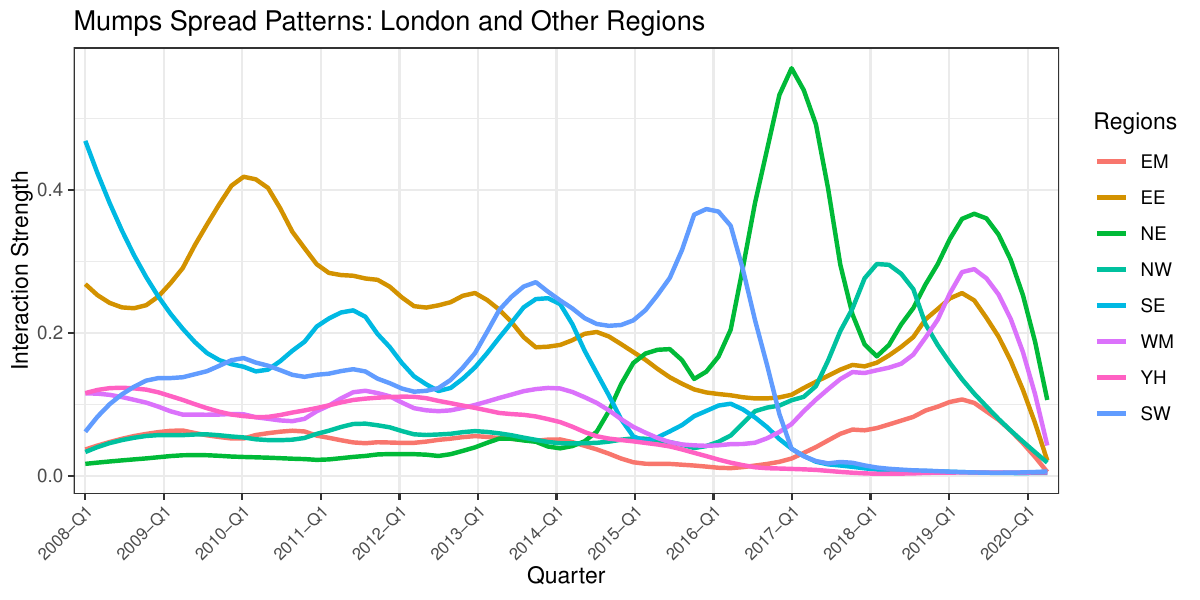}
    \caption{Smoothed temporal dynamics of interconnections between London and other regions.}
    \label{fig: smoothed_conn_london}
\end{figure}

London emerges as the most central node in the latent space overall, as depicted in Figure \ref{fig: dynamic_trajectory_mumps_cases}, illustrating its evolving connections with various regions over time. This centrality is further emphasized in Figure \ref{fig: smoothed_conn_london}, which shows how London's connections with other regions changes throughout the observed period. 
As the central node, London (LO) both influences and is influenced by different regions within the network. It consistently maintains a connection with its neighboring region, East of England (EE), throughout the entire time span. Furthermore, from 2015 onward, London (LO) develops significant links with the North East (NE) and North West (NW), particularly during periods when these regions experience peak mumps cases. This highlights London's crucial role in both driving and responding to mumps transmission dynamics across the network.

Figure \ref{fig: Latent space based connections in region} illustrates that the network estimated during the last peak period (2020 Q1) exhibits distinct clustering patterns among the regions. Specifically, regions such as London (LO), East Midlands (EM), East of England (EE), West Midlands (WM), and the North East (NE) form a cohesive cluster, indicating strong inter-regional connections and potential pathways for mumps transmission. In contrast, the South East (SE), South West (SW), and Yorkshire and Humberside (YH) are grouped into a separate cluster, highlighting their localized interaction in the spread of the infection. Notably, the North West (NW) region appears isolated, showing minimal connections with other regions during this period. This clustering pattern underscores the geographical dynamics in mumps transmission, where certain regions exhibit tightly knit connections, while others, like the North West (NW), seem to operate independently.

\subsection{Risk of contagion}

One major advantage of using LPMs is that they permit various useful model-based summaries to be created. These measures can be extracted as summaries of the latent space, thus being themselves dynamic over time. The empirical variance concept outlined by \textcite{Friel20166629} is a prime example of this. The measure enables us to quantify the level of expansion or contraction of the latent space. By measuring the empirical variance of the latent positions at each time, we can express numerically how close to each the nodes tend to be, at each point in time. This is a very valuable summary in this application, since it offers valuable insights into whether the risk of spread of infection strengthens or weakens over time. Indeed, a large empirical variance signals that the number of cases tend to evolve independently, suggesting that there is no spillover and contagion. By contrast, a contracted latent space is suggestive of strong interactions between the series and thus the possible presence of frequent contagion. 

\begin{figure}[htpb]
    \centering    
     \includegraphics[width = 0.8 \textwidth]{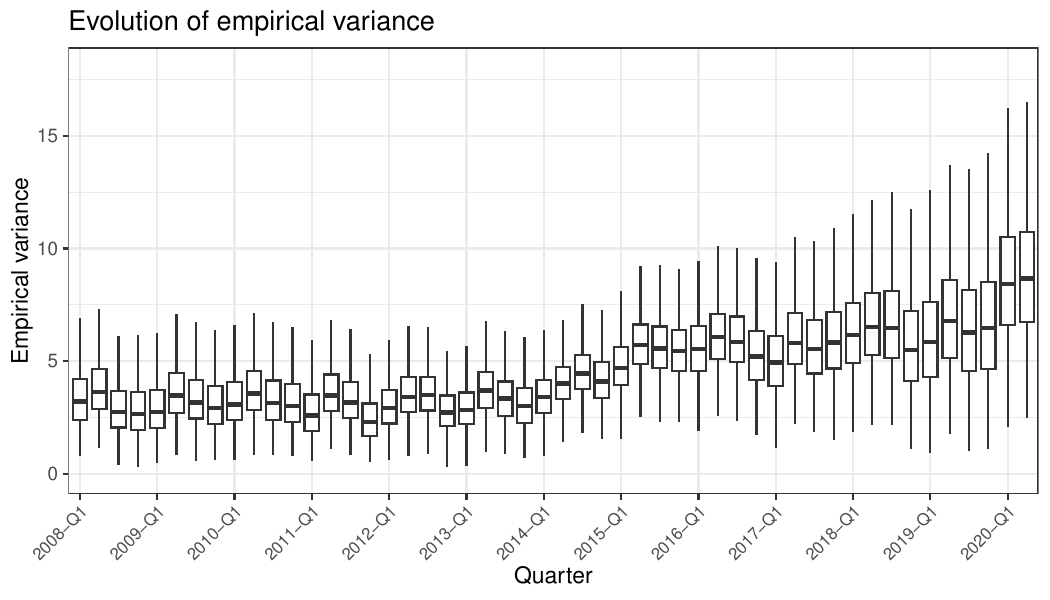}
    \caption{Temporal evolution of the posterior distribution of empirical variance for the period 2008 (Q1) to 2020 (Q2).}
    \label{fig: Emperical_varaince_Dy}
\end{figure}

The empirical variance is calculated from the output as follows:
\begin{equation}
\text{S}_{t} = \frac{1}{2N} \sum_{i=1}^{N} (\mathbf{z}_{it} - \Bar{\mathbf{z}}_{t})^{\top}
(\mathbf{z}_{it} - \Bar{\mathbf{z}}_{t})
\end{equation}
where $\Bar{\mathbf{z}}_{t}=\frac{\sum_{i=1}^{N} \mathbf{z}_{it}}{N}$ for $t=1,\dots ,T$ and ``$\top$'' represents the transpose. The above computation is performed for each HMC iteration. Figure \ref{fig: Emperical_varaince_Dy} illustrates the evolution of the empirical variance, displaying an upward trend over the observed period. 
This trend indicates that as time progresses, the interconnections between regions decrease, suggesting that the spread of mumps infection becomes less pronounced as the latent space expands and regions move farther apart from each other in the latent space. This pattern is also highlighted in Figure \ref{fig: Latent space based connections in region}, where edges between the regions seem to decrease as time progresses indicating lesser interconnection among regions. This growing distance between latent positions reflects a weakening of inter-regional connections, suggesting that the transmission of mumps cases among regions has become weaker over time.

\section{Conclusions} \label{discussion}
We have introduced a new methodology for the analysis of time series of counts, which relies on a multi-layer hierarchical model, whereby we infer both the interactions between the time series, and the network that generates those interactions.
The model has a clear advantage over other autoregressive time series models, in that it combines their flexibility with more clear parameter interpretations.
The network formulation allows us to create a parsimonious structure, which can help disentangle some of the complex interaction patterns that determine the observed datasets.

We have proposed an in-depth analysis of a dataset of mumps cases in England, using both a static and a dynamic model. Our objective was to estimate an underlying network of interactions between the regions of England, and how this network may change over time. An important aspect of our analysis is that we bypass the geographical information, to focus on a latent variable formulation that can directly capture other non-observed or non-observable factors. Our model provides a clear model based measure of the risk of contagion, which is directly created from the estimated latent space by measuring its expansions and contractions over time. The results from our model suggest that the level of contagion has been constantly decreasing over the last few years, possibly thanks to a higher uptake of the relevant vaccines. 
In addition, our results consistently highlight that the northern regions are the most vulnerable areas, and that the London area emerges as a central hub in both propagating and mitigating mumps transmission across the network. 

In this paper, we have proposed both a static and dynamic variants of our model, with results suggesting some relevant differences and use for them. The dynamic framework provides the most flexibility, but also a much higher number of model parameters. Through a number of simulations we have shown that either model can be accurately estimated on small datasets, thus recovering a variety of dependency patterns that we believe to be realistic.

Our work leaves a number of viable directions for future research. A main critical question that remains unanswered is definitely the computational scalability of our methodology. In this paper, we only focused on relatively small datasets, primarily due to long computing times. While STAN provides an accurate solution to the model estimation task, there are definitely other possible approaches that may be explored to speed up inference.

Secondly, a key question for latent position models regards the number of latent dimensions to choose. We did not pursue this research question here, and chose two dimensions as is common in many other works. However, selecting an ideal number of groups that is specific to the application can definitely lead to a superior methodology, especially considering that one main motivation for this paper was to address the over-parametrization that can occur in time series models.

Another comment can be made for the autoregressive order of a DTSLPM: in this paper we have only considered processes of order one, however, it could definitely be of interest to consider how the network structures may be used for processes of higher orders. Finally, differently from the recent work \textcite{Kaur_rastelli_2024}, we have not included covariates or seasonality in the DTSLPM formulation. However, we note that this may done in an analogous and straightforward way for those applications that would strongly require these features.

\section*{Acknowledgements}
This publication has emanated from research supported in part by a grant from the Insight Centre for Data Analytics which is supported by Science Foundation Ireland under Grant number $12/RC/2289\_P2$.

\printbibliography
\end{document}